\documentclass[iop]{emulateapj}

\usepackage{natbib}
\usepackage{ulem}
\usepackage{comment}
\usepackage{txfonts}
\usepackage{color}

\usepackage[T1]{fontenc}

\usepackage{bm}

\shorttitle{High-Velocity type Ia Supernova 2019ein}
\shortauthors{Kawabata M., et al.}

\begin{document}

\title{
Type Ia SN 2019ein: New Insights into the Similarities and diversities among High-Velocity SNe Ia
}
\author{
  Miho Kawabata\altaffilmark{1},
  Keiichi Maeda\altaffilmark{1},
  Masayuki Yamanaka\altaffilmark{2},
  Tatsuya Nakaoka\altaffilmark{3,4},
  Koji S. Kawabata\altaffilmark{3,4,5},
  Ryo Adachi\altaffilmark{6},
  Hiroshi Akitaya\altaffilmark{3,4},
  Umut Burgaz\altaffilmark{1,7}, 
  Hidekazu Hanayama\altaffilmark{8},
  Takashi Horiuchi\altaffilmark{8},
  Ryohei Hosokawa\altaffilmark{6},
  Kota Iida\altaffilmark{6},
  Fumiya Imazato\altaffilmark{3},  
  Keisuke Isogai\altaffilmark{2},
  Ji-an Jiang\altaffilmark{9},
  Noriyuki Katoh\altaffilmark{10},
  Hiroki Kimura\altaffilmark{3},
  Masaru Kino\altaffilmark{2},
  Daisuke Kuroda\altaffilmark{2},
  Hiroyuki Maehara\altaffilmark{11}, 
  Kazuya Matsubayashi\altaffilmark{2},
  Kumiko Morihana\altaffilmark{12},
  Katsuhiro L. Murata\altaffilmark{6},
  Takashi Nagao\altaffilmark{13},
  Masafumi Niwano\altaffilmark{6},
  Daisaku Nogami\altaffilmark{1},
  Motoki Oeda\altaffilmark{6},
  Tatsuharu Ono\altaffilmark{14},
  Hiroki Onozato\altaffilmark{10},
  Masaaki Otsuka\altaffilmark{2},
  Tomoki Saito\altaffilmark{10},
  Mahito Sasada\altaffilmark{3,4,5},
  Kazuki Shiraishi\altaffilmark{6},
  Haruki Sugiyama\altaffilmark{14},
  Kenta Taguchi\altaffilmark{1},
  Jun Takahashi\altaffilmark{10},
  Kengo Takagi\altaffilmark{3},
  Seiko Takagi\altaffilmark{14},
  Masaki Takayama\altaffilmark{10},
  Miyako Tozuka\altaffilmark{10}, and 
  Kazuhiro Sekiguchi\altaffilmark{15}
}

\altaffiltext{1}{Department of Astronomy, Kyoto University, Kitashirakawa-Oiwakecho, Sakyo-ku, Kyoto 606-8502, Japan; kawabata@kusastro.kyoto-u.ac.jp}
\altaffiltext{2}{Okayama Observatory, Kyoto University, 3037-5 Honjo, Kamogata-cho, Asakuchi, Okayama 719-0232, Japan}
\altaffiltext{3}{Hiroshima Astrophysical Science Center, Hiroshima University, Higashi-Hiroshima, Hiroshima 739-8526, Japan}
\altaffiltext{4}{Department of Physical Science, Hiroshima University, Kagamiyama 1-3-1, Higashi-Hiroshima 739-8526, Japan}
\altaffiltext{5}{Core Research for Energetic Universe (CORE-U), Hiroshima University, Kagamiyama, Higashi-Hiroshima, Hiroshima 739-8526, Japan}
\altaffiltext{6}{Department of Physics, Tokyo Institute of Technology, 2-12-1 Ookayama, Meguro-ku, Tokyo 152-8551, Japan}
\altaffiltext{7}{Department of Astronomy and Space Sciences, Ege University, 35100 Izmir, Turkey}
\altaffiltext{8}{Ishigakijima Astronomical Observatory, National Astronomical Observatory of Japan, 1024-1 Arakawa, Ishigaki, Okinawa 907-0024, Japan}
\altaffiltext{9}{Kavli Institute for the Physics and Mathematics of the Universe (WPI), The University of Tokyo Institutes for Advanced Study, The University of Tokyo, 5-1-5 Kashiwanoha, Kashiwa, Chiba 277-8583, Japan}
\altaffiltext{10}{Nishi-Harima Astronomical Observatory, Center for Astronomy, University of Hyogo, 407-2 Nishigaichi, Sayo-cho, Sayo, Hyogo 679-5313, Japan}
\altaffiltext{11}{Okayama Branch Office, Subaru Telescope, National Astronomical Observatory of Japan, NINS, Kamogata, Asakuchi, Okayama 719-0232, Japan}
\altaffiltext{12}{Graduate School of Science, Nagoya University, Furo-cho, Chikusa-ku, Nagoya 464-8602, Japan}
\altaffiltext{13}{European Southern Observatory, Karl-Schwarzschild-Str. 2, 85748 Garching b. M\"{u}nchen, Germany}
\altaffiltext{14}{Department of Cosmosciences, Graduate School of Science, Hokkaido University, Kita 10 Nishi8, Kita-ku, Sapporo 060-0810, Japan}
\altaffiltext{15}{National Astronomical Observatory of Japan, 2-21-1 Osawa, Mitaka, Tokyo 181-8588, Japan}

\begin{abstract}
We present optical observations of type Ia supernova (SN) 2019ein, 
starting at 2 days after the estimated explosion date. 
The spectra and the light curves show that SN 2019ein belongs to the 
High-Velocity (HV) and Bload Line groups with relatively 
rapid decline in the light curves ($\Delta m_{15}(B) = 1.36 \pm 0.02$ mag) 
and the short rise time (15.37 $\pm$ 0.55 days).
The Si {\sc ii} $\lambda$6355 velocity, associated with a photospheric 
component but not with a detached high-velocity feature, reached 
$\sim$ 20,000 km s$^{-1}$ at $12$ days before the $B$-band maximum. 
The line velocity however decreased very rapidly and smoothly toward the 
maximum light, where it was $\sim 13,000$ km s$^{-1}$ as relatively low 
among HV SNe.
This indicates that the speed of the spectral evolution of HV SNe Ia is 
correlated not only to the velocity at 
the maximum light, but also to the light curve decline rate like the 
case for Normal-Velocity (NV) SNe Ia.
Spectral synthesis modeling shows that the outermost layer at 
$> 17,000$ km s$^{-1}$ is well described by the O-Ne-C burning layer 
extending to at least $25,000$ km s$^{-1}$, and there is no unburnt carbon 
below $30,000$ km s$^{-1}$; these properties are largely consistent with 
the delayed detonation scenario, and are shared with the prototypical HV 
SN 2002bo despite the large difference in $\Delta m_{15}(B)$. 
This structure is strikingly different from that derived for the 
well-studied NV SN 2011fe. 
We suggest that the relation between the mass of $^{56}$Ni 
(or $\Delta m_{15}$) and the extent of the O-Ne-C burning layer provides 
an important constraint on the explosion mechanism(s) of HV and NV SNe. 
\end{abstract}


\section{Introduction}\label{sec:intro}

It has been widely accepted that type Ia supernovae (SNe Ia) arise 
from the explosion of a white dwarf (WD) in a binary system. 
When the WD mass reaches nearly to the Chandrasekhar-limiting mass, 
the thermonuclear runaway is expected to be triggered.
For SNe Ia, there is a well-established correlation between the peak 
luminosity and the light-curve decline rate, which is known as the 
luminosity-width relation (\citealt{Phillips93}). 
This relation allows SNe Ia to be used as precise standardized candles 
to measure the cosmic-scale distance to remote galaxies and thus the 
cosmological parameters (\citealt{Riess98}; \citealt{Perlmutter99}).
Despite the importance of SNe Ia both in cosmology and astrophysics, 
the progenitor(s), the explosion mechanism(s) and the origin(s) of 
diversities are still under active debate. 

SNe Ia show the spectroscopic diversity (e.g., \citealt{Benetti05}; 
\citealt{Branch06}; \citealt{Wang09}; \citealt{Blondin12}).
\citet{Benetti05} found that SNe Ia are classified into three different 
groups in terms of the velocity of the Si {\sc ii} $\lambda$6355 line 
around the maximum light. 
(1) The first group has a small expansion velocity, and the evolution 
(decrease) of the velocity is fast. 
This group consists of SN 1991bg-like faint SNe Ia. 
(2) The second group is the low velocity gradient (LVG) group. 
SNe belonging to this group have a low (or normal) expansion velocity, 
and slow evolution in the line velocity decrease. 
(3) The third group has a high expansion velocity, and the velocity 
decreases rapidly with time. It is called the high velocity gradient 
(HVG) group. 
The HVG group contains photometrically normal SNe Ia, e.g., SNe 2002bo 
(\citealt{Benetti04}), 2002dj (\citealt{Pignata04}) and 
2006X (\citealt{Wang08}; \citealt{Yamanaka09}).

SNe Ia classified into the HVG group generally have larger 
velocities than those in the LVG group. 
\citet{Wang09} introduced a different classification, using the mean 
Si {\sc ii} $\lambda$6355 velocity within one week since the $B$-band 
maximum. 
They defined the average Si {\sc ii} velocity using 10 well-observed 
normal SNe Ia, and then classified SNe Ia that have a significantly 
high velocity beyond a 3 $\sigma$ level as 
``High-Velocity'' (HV) SNe Ia. 
The remaining SNe Ia are classified as ``Normal-Velocity'' SNe Ia (NV 
SNe Ia; but see \citealt{Blondin12}). 
The dividing velocity between HV and NV SNe is 
$\sim -12,000$ km s$^{-1}$ at the maximum light. 
We will follow this terminology in this work unless otherwise mentioned. 

It has been suggested that HV SNe show different properties from 
NV SNe beside the line velocity and its evolution. 
HV SNe have a redder intrinsic $B-V$ color 
(e.g., \citealt{Pignata08}). 
They also seem to have a different extinction law from NV SNe 
(e.g., \citealt{Wang09}; \citealt{Foley11b}); the extinction 
parameter, $R_{V} = A_{V}/E(B-V)$, is generally lower than the 
standard Milky Way (MW) value. 
The emission line shifts in the late phase are also different between 
HV and NV SNe (\citealt{Maeda10}). 
\citet{Ganeshalingam11} suggested that HV SNe have a shorter rise 
time (the time interval between the explosion and the maximum light) 
than NV SNe.

The origin of these spectroscopic subclasses has not been fully clarified. 
Furthermore, it is not yet clear whether they form distinct groups 
or one continuum group (e.g., \citealt{Benetti05}).
It has been pointed out that (some of) these properties 
might be explained by an (intrinsically same) asymmetric explosion but 
viewed from different directions (e.g., \citealt{Maeda10}; 
\citealt{Maeda11}). 
There are also observational indications for intrinsically different 
populations between the HV and NV SNe groups (\citealt{Wang13}; 
\citealt{Wang18}), adding further compilations on their origins. 
We note that these two suggestions might indeed not be mutually 
exclusive; for example, there could indeed be two populations within 
the NV SNe group, where one is overlapping with the HV SNe group but 
the other is not. 

Our understanding of the nature of HV SNe and its difference to 
NV SNe is still far from satisfactory. 
In the last decade, there has been an increasing attention to the 
importance of the observational data starting within a few days since 
the explosion, in extending our understanding on the natures of the 
progenitor(s) and explosion mechanism(s) of SNe Ia. 
However, such data are very limited for HV SNe. 
This is a main issue we deal with in this work. 

The light curve behavior within the first few days since the explosion 
provides powerful diagnostics on the nature of the progenitor system. 
If the SN ejecta collide with an extended, non-degenerate companion 
star, it is expected to create a detectable signature as an excessive 
emission in the first few days, if the companion star is sufficiently 
large in its radius (e.g., \citealt{Kasen10}; \citealt{Kutsuna15}; 
\citealt{Maeda18}). 
Mostly for NV SNe, the approach has been used to constrain the nature 
of the possible companion star.
As an example, the observations of SN 2011fe within one day since the 
explosion provided a strong constraint ($\leq 0.1 - 0.25 R_{\odot}$; 
\citealt{Nugent11}, \citealt{Bloom12}, \citealt{Goobar15}).
In the last decade, there are an increasing number of NV SNe for which 
similar constraints have been placed (e.g., \citealt{Bianco11}; 
\citealt{Foley12}; \citealt{Silverman12}; \citealt{Zheng13}; 
\citealt{Yamanaka14}; \citealt{Goobar15}; \citealt{Im15}; 
\citealt{Olling15}; \citealt{Marion16}; \citealt{Shappee16}; 
\citealt{Hosseinzadeh17}; \citealt{Contreras18}; \citealt{Holmbo18}; 
\citealt{Miller18}; \citealt{Shappee18}; \citealt{Dimitriadis19}; 
\citealt{Shappee19}).

At the same time, diverse nature of the earliest light curve 
properties has also been noticed. 
The light curves in the early phase are typically 
well-fitted by a single power-law function 
(e.g., \citealt{Nugent11}).
However, some SNe Ia showing noticeable deviation from this behavior 
have also been discovered (e.g., \citealt{Hosseinzadeh17}; 
\citealt{Contreras18}; \citealt{Miller18}; \citealt{Jiang17}; 
\citealt{Jiang20}), which require a fit by a broken power-law function 
(\citealt{Zheng13}; \citealt{Zheng14}). 
There is also a suggestion that the bright SN 1991T/1999aa-like SNe 
tend to show the excessive emission in the first few days (e.g., 
\citealt{Stritzinger18}; \citealt{Jiang18}), 
but it is probably caused by a mechanism other than the companion 
interaction (\citealt{Maeda18}). 

Spectroscopic data in the earliest phase shortly after the 
explosion are useful in refining our understanding on 
the nature of SNe Ia. 
The nature of the outermost layer can be studied using the spectra 
within a week after the explosion (\citealt{Stehle05}). 
In addition, if the spectra are taken at sufficiently 
early epochs, most of SNe Ia show the so-called High-Velocity Features 
(HVFs) in some spectral lines, which seem to exist separately from the 
photospheric component\footnote{We call the (lower-velocity) main 
absorption component as a photospheric component following convention, 
while it does not have to be formed exactly at the photosphere.} 
(e.g., \citealt{Mazzali01}; \citealt{Mattila05}; 
\citealt{Mazzali05a}; \citealt{Garavini07}; \citealt{Stanishev07}; 
\citealt{Childress13}; \citealt{Zhao16}).
The origin of the HVFs is yet to be clarified (e.g., 
\citealt{Gerardy04}; \citealt{Mazzali05a}; \citealt{Tanaka06}; 
\citealt{Tanaka08}), but it likely contains information beyond our 
current understanding on the progenitor and explosion mechanism of 
SNe Ia. 

The observational data starting within a few days since the explosion 
are extremely rare for HV SNe. 
The best-studied HV SNe 2002bo and 2006X have the spectroscopic data 
only after a week since the explosion. 
Both of them show relatively slow evolution in their light curves. 
The best studied HV SN with relatively fast evolution in the light 
curves is SN 2002er, but again the data in the earliest phase are 
missing. 
Furthermore, in the classification scheme of \citet{Wang09}, 
SN 2002er is a transitional object between HV and NV groups, and it is 
thus not the best object to study the difference between the two groups.

Therefore, it is very important to obtain the observational data 
starting within a few days since the explosion for a HV SN, 
especially for those showing rapid light curve evolution. 
It will then provide various diagnostics on the nature of the 
progenitor and the explosion mechanism as mentioned 
above, and it will allow to identify possible 
diversities within the HV SN group and investigate possible 
relations between NV and HV SNe. 

SN 2019ein was discovered at 18.194 mag in NGC5353 on 2019 May 1.5 UT 
by the Asteroid Terrestrial-impact Last Alert System (ATLAS) project 
(ATLAS19ieo; \citealt{Tonry19}).
The spectrum was obtained on May 2.3 UT by the Las Cumbres Observatory 
(LCO) Global SN project, and this SN was 
classified as a SN Ia (\citealt{Burke19}) at about two weeks before 
the maximum light.
They reported that the best-fit SNe Ia contain HV SN 2002bo. 
The spectral features showed the high velocity and blended 
Ca {\sc ii} IR triplet and O {\sc i} lines. 
In this paper, we report the multi-band observations of SN 2019ein. 
We describe the observation and data reduction in \S\ref{sec:obs}.
We present the results of the observations, and compare its 
properties with well-studied HV SN 2002bo and other HV SNe Ia in 
\S\ref{sec:results}. 
In \S\ref{sec:disc}, we discuss the nature of the progenitor and 
the explosion mechanism of SN 2019ein based on the photometric and 
spectral data obtained within a few days from the explosion.
A summary of this work is provided in \S\ref{sec:conclusion}.

\section{Observations and Data Reduction}\label{sec:obs}
We performed spectral observations of SN 2019ein using HOWPol 
mounted on the Kanata telescope of Hiroshima Unviersity, and 
KOOLS-IFU on the Seimei telescope of Kyoto University. 
Multi-band imaging observations were conducted as a 
Target-of-Opportunity (ToO) program in the framework of the 
Optical and Infrared Synergetic Telescopes for Education and 
Research (OISTER).
All the magnitudes from UV to NIR are given 
in the Vega system.

\subsection{Photometry}\label{sec:photo}
We performed $BVRI$-band imaging observations using Hiroshima One-shot 
Wide-field Polarimeter (HOWPol; \citealt{Kawabata08}) installed on the 
Nasmyth stage of the 1.5-m Kanata telescope at the Higashi-Hiroshima 
Observatory, Hiroshima University. 
$UBVRI$-band images were also taken using the Multi-spectral Imager 
(MSI; \citealt{Watanabe12}) installed to the 1.6-m Pirka telescope. 
Additionally, $g'RI$-band imaging observations were performed using 
a robotic observation system with the Multicolor Imaging Telescopes 
for Survey and Monstrous Explosions (MITSuME; \citealt{Kotani05}) at 
the Akeno Observatory (AO) and at the Ishigaki-jima Astronomical 
Observatory (IAO).

\begin{figure}[t]
\centering
\includegraphics[width=8.0cm,clip]{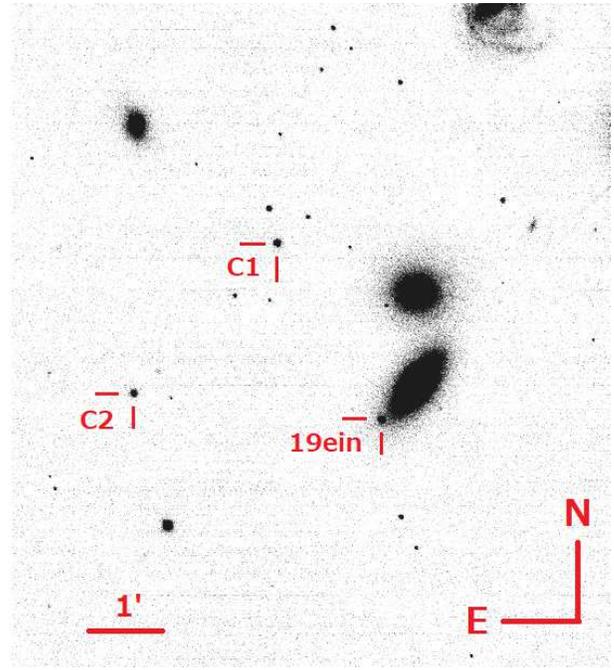}
\caption{$R$-band image of SN 2019ein and the comparison stars taken 
with the Kanata telescope / HOWPol on MJD 58615.7 (2019 May 12).}
\label{fig:cmp_star}
\end{figure}

We reduced the imaging data in a standard manner for the CCD 
photometry. 
We adopted the Point-Spread-Function (PSF) fitting photometry 
method using DAOPHOT package in {\it IRAF}\footnote{{\it IRAF} 
is distributed by the National Optical Astronomy Observatory, which 
is operated by the Association of Universities for Research in 
Astronomy (AURA) under a cooperative agreement with the National 
Science Foundation.}, without subtracting the galaxy template.
For the data in the first few days, we have checked the contamination 
from the host galaxy by subtracting the SDSS images 
from our images. 
Within the uncertainty set by the difference in the photometric 
filters, the result of this exercise agrees with the PSF photometry. 
Therefore, it would not significantly affect the main conclusions 
in the present work. 
We either skipped the S-correction, since it is negligible for 
the purposes of the present study (\citealt{Stritzinger02}).
For the magnitude calibration, we adopted relative photometry using the 
comparison stars taken in the same frames (Figure \ref{fig:cmp_star}). 
The magnitudes of the comparison stars in the $UBVRI$ bands were 
calibrated with the stars in the M100 field (\citealt{Wang08}) 
observed on a photometric night, as shown in Table \ref{tb:cmp_star}.
The magnitude in the $g'$ band was calibrated 
using the SDSS DR12 photometric catalog (\citealt{Alam15}).
First-order color term correction was applied in the photometry. 
All optical photometric results are listed in Table \ref{tb:opt_log}

\begin{deluxetable*}{cccccc}
\tablewidth{0pt}
\tablecaption{Magnitudes of the comparison stars.}
\tablehead{
 ID &  $U$  &  $B$  &  $V$  &  $R$  & $I$ \\
    & (mag) & (mag) & (mag) & (mag) & (mag)
}
\startdata
C1 & 14.780 $\pm$ 0.099 & 14.621 $\pm$ 0.013 & 14.243 $\pm$ 0.012 & 14.011 $\pm$ 0.017 & 13.619 $\pm$ 0.023\\
C2 & 15.583 $\pm$ 0.116 & 15.037 $\pm$ 0.015 & 14.478 $\pm$ 0.014 & 14.120 $\pm$ 0.019 & 13.638 $\pm$ 0.023
\enddata
\label{tb:cmp_star}
\end{deluxetable*}

\begin{deluxetable*}{cccccccccc} 
\tablewidth{0pt}
\tablecaption{Log of optical observations of SN 2019ein.}
\tablehead{
  Date  & MJD  &  Phase\tablenotemark{a}  & $U$   & $B$   & $g'$  &  $V$   & $R$   &  $I$  & Telescope \\
        &      &  (day)  & (mag) & (mag) & (mag) &  (mag) & (mag) & (mag) & (Instrument)  
}
\startdata
2019-05-01 & 58604.8 & $-13.4$ &         --         & 18.255 $\pm$ 0.044 &         --         & 18.010 $\pm$ 0.077 & 17.619 $\pm$ 0.140 & 17.869 $\pm$ 0.023 & Kanata (HOWPol) \\
2019-05-02 & 58605.5 & $-12.7$ &         --         &         --         &         --         & 17.211 $\pm$ 0.044 & 17.059 $\pm$ 0.103 & 17.318 $\pm$ 0.100 & Kanata (HOWPol) \\
2019-05-03 & 58606.8 & $-11.5$ &         --         & 16.549 $\pm$ 0.019 &         --         & 16.473 $\pm$ 0.013 & 16.294 $\pm$ 0.018 & 16.284 $\pm$ 0.024 & Kanata (HOWPol) \\
2019-05-04 & 58607.6 & $-10.7$ &         --         &         --         &         --         & 16.140 $\pm$ 0.012 &         --         & 15.852 $\pm$ 0.023 & Kanata (HOWPol) \\
2019-05-05 & 58608.5 &  $-9.8$ &         --         &         --         & 15.454 $\pm$ 0.066 &         --         &         --         &         --         & (MITSuME/AO)    \\
2019-05-06 & 58609.7 &  $-8.6$ &         --         & 15.231 $\pm$ 0.018 &         --         & 15.260 $\pm$ 0.014 & 15.090 $\pm$ 0.017 & 15.022 $\pm$ 0.023 & Kanata (HOWPol) \\
2019-05-07 & 58610.5 &  $-7.8$ &         --         &         --         & 14.845 $\pm$ 0.049 &         --         &         --         &         --         & (MITSuME/AO)    \\
2019-05-07 & 58610.7 &  $-7.6$ &         --         & 14.924 $\pm$ 0.014 &         --         & 14.967 $\pm$ 0.012 & 14.823 $\pm$ 0.018 & 14.775 $\pm$ 0.023 & Kanata (HOWPol) \\
2019-05-08 & 58611.5 &  $-6.8$ &         --         &         --         & 14.737 $\pm$ 0.031 &         --         &         --         &         --         & (MITSuME/AO)    \\
2019-05-08 & 58611.5 &  $-6.7$ & 14.301 $\pm$ 0.207 & 14.635 $\pm$ 0.014 &         --         & 14.569 $\pm$ 0.191 & 14.657 $\pm$ 0.018 & 14.524 $\pm$ 0.028 & Pirika (MSI)    \\
2019-05-09 & 58612.5 &  $-5.8$ &         --         &         --         & 14.625 $\pm$ 0.003 &         --         &         --         &         --         & (MITSuME/AO)    \\
2019-05-09 & 58612.6 &  $-5.6$ &         --         & 14.446 $\pm$ 0.014 &         --         & 14.506 $\pm$ 0.013 & 14.418 $\pm$ 0.017 & 14.399 $\pm$ 0.025 & Kanata (HOWPol) \\
2019-05-11 & 58614.5 &  $-3.7$ & 14.036 $\pm$ 0.408 & 14.146 $\pm$ 0.015 &         --         & 14.175 $\pm$ 0.016 & 14.223 $\pm$ 0.017 & 14.185 $\pm$ 0.024 & Pirika (MSI)    \\
2019-05-11 & 58614.5 &  $-3.7$ &         --         &         --         & 14.287 $\pm$ 0.023 &         --         &         --         & 14.106 $\pm$ 0.039 & (MITSuME/IAO)   \\
2019-05-11 & 58614.7 &  $-3.6$ &         --         & 14.178 $\pm$ 0.014 &         --         & 14.205 $\pm$ 0.015 & 14.218 $\pm$ 0.019 & 14.218 $\pm$ 0.026 & Kanata (HOWPol) \\
2019-05-12 & 58615.6 &  $-2.6$ & 13.703 $\pm$ 0.107 &         --         &         --         & 14.072 $\pm$ 0.013 & 14.116 $\pm$ 0.018 & 14.084 $\pm$ 0.024 & Pirika (MSI)    \\
2019-05-12 & 58615.7 &  $-2.5$ &         --         & 14.104 $\pm$ 0.013 &         --         & 14.104 $\pm$ 0.015 & 14.160 $\pm$ 0.017 & 14.186 $\pm$ 0.025 & Kanata (HOWPol) \\
2019-05-13 & 58616.5 &  $-1.7$ &         --         & 14.005 $\pm$ 0.014 &         --         & 14.017 $\pm$ 0.013 & 14.098 $\pm$ 0.017 & 14.134 $\pm$ 0.024 & Pirika (MSI)    \\
2019-05-14 & 58617.6 &  $-0.6$ & 13.600 $\pm$ 0.100 & 13.972 $\pm$ 0.013 &         --         & 13.952 $\pm$ 0.013 & 14.061 $\pm$ 0.017 & 14.151 $\pm$ 0.029 & Pirika (MSI)    \\
2019-05-14 & 58617.6 &  $-0.6$ &         --         &         --         & 13.986 $\pm$ 0.045 &         --         &         --         &         --         & (MITSuME/IAO)   \\
2019-05-15 & 58618.6 &    0.4  & 13.687 $\pm$ 0.099 & 13.977 $\pm$ 0.013 &         --         & 13.942 $\pm$ 0.013 & 14.050 $\pm$ 0.017 & 14.179 $\pm$ 0.023 & Pirika (MSI)    \\
2019-05-15 & 58618.7 &    0.4  &         --         & 14.013 $\pm$ 0.013 &         --         & 13.939 $\pm$ 0.014 & 14.072 $\pm$ 0.019 & 14.238 $\pm$ 0.024 & Kanata (HOWPol) \\
2019-05-16 & 58619.5 &    1.2  &         --         &         --         & 14.122 $\pm$ 0.015 &         --         &         --         &         --         & (MITSuME/AO)    \\
2019-05-16 & 58619.6 &    1.4  &         --         & 14.058 $\pm$ 0.020 &         --         & 14.152 $\pm$ 0.113 &         --         & 14.279 $\pm$ 0.060 & Pirika (MSI)    \\
2019-05-17 & 58620.6 &    2.4  &         --         & 14.051 $\pm$ 0.014 &         --         & 14.118 $\pm$ 0.140 &         --         & 14.289 $\pm$ 0.023 & Pirika (MSI)    \\
2019-05-18 & 58621.6 &    3.4  & 13.769 $\pm$ 0.099 & 14.146 $\pm$ 0.013 &         --         &         --         & 14.080 $\pm$ 0.017 & 14.284 $\pm$ 0.024 & Pirika (MSI)    \\
2019-05-19 & 58622.5 &    4.3  &         --         &         --         & 14.140 $\pm$ 0.018 &         --         &         --         &         --         & (MITSuME/AO)    \\
2019-05-19 & 58622.6 &    4.4  & 13.915 $\pm$ 0.103 & 14.110 $\pm$ 0.014 &         --         & 14.016 $\pm$ 0.114 & 14.091 $\pm$ 0.017 & 14.288 $\pm$ 0.025 & Pirika (MSI)    \\
2019-05-21 & 58624.6 &    6.4  &         --         &         --         & 14.187 $\pm$ 0.012 &         --         &         --         & 14.259 $\pm$ 0.028 & (MITSuME/IAO)   \\
2019-05-23 & 58626.5 &    8.2  &         --         &         --         & 14.349 $\pm$ 0.027 &         --         &         --         &         --         & (MITSuME/AO)    \\
2019-05-23 & 58626.6 &    8.4  &         --         & 14.516 $\pm$ 0.017 &         --         & 14.126 $\pm$ 0.012 & 14.446 $\pm$ 0.017 & 14.689 $\pm$ 0.024 & Kanata (HOWPol) \\
2019-05-24 & 58627.5 &    9.3  &         --         &         --         & 14.413 $\pm$ 0.031 &         --         &         --         &         --         & (MITSuME/AO)    \\
2019-05-25 & 58628.5 &   10.2  &         --         & 14.731 $\pm$ 0.013 &         --         & 14.276 $\pm$ 0.012 & 14.599 $\pm$ 0.020 & 14.822 $\pm$ 0.025 & Kanata (HOWPol) \\
2019-05-25 & 58628.6 &   10.3  &         --         &         --         & 14.427 $\pm$ 0.025 &         --         &         --         & 14.613 $\pm$ 0.032 & (MITSuME/IAO)   \\
2019-05-26 & 58629.5 &   11.2  &         --         &         --         & 14.585 $\pm$ 0.035 &         --         &         --         &         --         & (MITSuME/AO)    \\
2019-05-26 & 58629.6 &   11.4  &         --         & 14.817 $\pm$ 0.013 &         --         & 14.372 $\pm$ 0.013 & 14.682 $\pm$ 0.019 & 14.871 $\pm$ 0.024 & Kanata (HOWPol) \\
2019-05-29 & 58632.6 &   14.4  &         --         & 15.274 $\pm$ 0.014 &         --         & 14.586 $\pm$ 0.012 & 14.742 $\pm$ 0.019 & 14.808 $\pm$ 0.023 & Kanata (HOWPol) \\
2019-06-03 & 58637.6 &   19.3  &         --         & 15.935 $\pm$ 0.018 &         --         & 14.842 $\pm$ 0.012 & 14.747 $\pm$ 0.018 & 14.612 $\pm$ 0.025 & Kanata (HOWPol) \\
2019-06-08 & 58642.6 &   24.4  &         --         & 16.515 $\pm$ 0.013 &         --         & 15.181 $\pm$ 0.014 & 15.009 $\pm$ 0.023 & 14.625 $\pm$ 0.026 & Kanata (HOWPol) \\
2019-06-12 & 58646.6 &   28.3  &         --         & 16.879 $\pm$ 0.020 &         --         & 15.526 $\pm$ 0.015 & 15.386 $\pm$ 0.020 & 14.994 $\pm$ 0.028 & Kanata (HOWPol) \\
2019-06-15 & 58649.6 &   31.4  &         --         &         --         & 16.598 $\pm$ 0.019 &         --         &         --         & 15.046 $\pm$ 0.032 & (MITSuME/IAO)   \\
2019-06-17 & 58651.6 &   33.4  &         --         &         --         &         --         & 15.775 $\pm$ 0.013 & 15.722 $\pm$ 0.018 & 15.364 $\pm$ 0.023 & Kanata (HOWPol) \\
2019-06-18 & 58652.5 &   34.3  &         --         &         --         &         --         &         --         &         --         &         --         & (MITSuME/AO)    \\
2019-06-19 & 58653.5 &   35.3  &         --         & 17.182 $\pm$ 0.021 &         --         & 15.868 $\pm$ 0.020 & 15.841 $\pm$ 0.020 & 15.493 $\pm$ 0.026 & Kanata (HOWPol) \\
2019-06-23 & 58657.5 &   39.3  &         --         & 17.250 $\pm$ 0.017 &         --         & 15.978 $\pm$ 0.013 & 15.978 $\pm$ 0.020 & 15.697 $\pm$ 0.026 & Kanata (HOWPol) \\
2019-06-25 & 58659.6 &   41.4  &         --         & 17.239 $\pm$ 0.021 &         --         & 16.029 $\pm$ 0.014 & 16.058 $\pm$ 0.018 & 15.791 $\pm$ 0.026 & Kanata (HOWPol) \\
2019-07-02 & 58666.5 &   48.3  &         --         & 17.327 $\pm$ 0.110 &         --         & 16.230 $\pm$ 0.015 & 16.329 $\pm$ 0.030 & 16.118 $\pm$ 0.035 & Kanata (HOWPol) \\
2019-07-04 & 58668.5 &   50.3  &         --         & 17.327 $\pm$ 0.021 &         --         & 16.282 $\pm$ 0.014 & 16.392 $\pm$ 0.018 & 16.242 $\pm$ 0.024 & Kanata (HOWPol) 
\enddata
\label{tb:opt_log}
\tablenotetext{a}{Relative to the epoch of $B$-band maximum (MJD 58618.24).}
\end{deluxetable*}

We also performed $JHK_{\rm s}$-band imaging observations with the 
Hiroshima Optical and Near-InfraRed camera (HONIR; \citealt{Akitaya14}) 
attached to the 1.5-m Kanata telescope, and with the Nishi-harima 
Infrared Camera (NIC) installed at the Cassegrain focus of the 2.0-m 
Nayuta telescope at 
the NishiHarima Astronomical Observatory. 
We adopted the sky background subtraction using a template sky image 
obtained by the dithering observation. 
We performed the PSF fitting photometry in the same way as used for 
the reduction of the optical data, and calibrated the 
magnitudes using the comparison stars in the 2MASS 
catalog (\citealt{Persson98}). 
All NIR photometric results are listed in Table \ref{tb:nir_log}.

\begin{deluxetable*}{ccccccc} 
\tablewidth{0pt}
\tablecaption{Log of NIR observations of SN 2019ein.}
\tablehead{
  Date  & MJD  &  Phase\tablenotemark{b}  & $J$   & $H$   & $K_{\rm s}$ & Telescope \\
        &      &  (day)  & (mag) & (mag) & (mag)       & (Instrument)  
}
\startdata
2019-05-03 & 58606.7 & $-11.5$ & 16.315 $\pm$ 0.030 & 16.085 $\pm$ 0.120 &          --        & Kanata (HONIR) \\
2019-05-05 & 58608.7 &  $-9.5$ &         --         & 15.741 $\pm$ 0.031 &          --        & Kanata (HONIR) \\
2019-05-06 & 58609.7 &  $-8.5$ & 14.955 $\pm$ 0.033 & 14.821 $\pm$ 0.079 & 14.984 $\pm$ 0.067 & Kanata (HONIR) \\
2019-05-07 & 58610.7 &  $-7.5$ & 14.792 $\pm$ 0.032 & 14.801 $\pm$ 0.050 &	15.108 $\pm$ 0.042 & Kanata (HONIR) \\
2019-05-09 & 58612.7 &  $-5.5$ & 14.597 $\pm$ 0.121 & 14.877 $\pm$ 0.223 &	14.578 $\pm$ 0.131 & Kanata (HONIR) \\
2019-05-10 & 58613.7 &  $-4.5$ & 14.610 $\pm$ 0.033 & 14.687 $\pm$ 0.074 &	14.598 $\pm$ 0.166 & Kanata (HONIR) \\
2019-05-12 & 58615.7 &  $-2.5$ & 14.515 $\pm$ 0.050 &         --         & 14.476 $\pm$ 0.078 & Kanata (HONIR) \\
2019-05-14 & 58617.5 &  $-0.7$ & 14.640 $\pm$ 0.124 & 14.803 $\pm$ 0.083 & 14.490 $\pm$ 0.139 & Nayuta (NIC)   \\
2019-05-14 & 58617.6 &  $-0.6$ & 14.747 $\pm$ 0.033 & 14.823 $\pm$ 0.223 &	14.586 $\pm$ 0.069 & Kanata (HONIR) \\
2019-05-15 & 58618.6 &    0.4  & 14.510 $\pm$ 0.019 & 14.808 $\pm$ 0.109 &          --        & Kanata (HONIR) \\
2019-05-15 & 58618.7 &    0.5  & 14.457 $\pm$ 0.070 & 14.846 $\pm$ 0.097 & 14.503 $\pm$ 0.195 & Nayuta (NIC)   \\ 
2019-05-16 & 58619.6 &    1.4  & 14.549 $\pm$ 0.040 & 14.770 $\pm$ 0.140 &          --        & Kanata (HONIR) \\
2019-05-16 & 58619.7 &    1.5  & 14.871 $\pm$ 0.088 & 14.889 $\pm$ 0.116 & 14.560 $\pm$ 0.189 & Nayuta (NIC)   \\ 
2019-05-21 & 58624.5 &    6.3  &         --         & 15.047 $\pm$ 0.117 & 14.797 $\pm$ 0.209 & Nayuta (NIC)   \\ 
2019-05-21 & 58624.6 &    6.4  & 15.186 $\pm$ 0.025 & 14.976 $\pm$ 0.051 &	14.615 $\pm$ 0.027 & Kanata (HONIR) \\
2019-05-22 & 58625.6 &    7.4  & 15.484 $\pm$ 0.035 & 15.370 $\pm$ 0.033 &	14.809 $\pm$ 0.045 & Kanata (HONIR) \\
2019-05-24 & 58627.7 &    9.5  &         --         &          --        & 14.896 $\pm$ 0.221 & Nayuta (NIC)   \\
2019-05-28 & 58631.7 &   13.5  &         --         & 14.938 $\pm$ 0.039 &          --        & Kanata (HONIR) \\
2019-05-29 & 58632.5 &   14.3  &         --         & 15.051 $\pm$ 0.115 & 14.944 $\pm$ 0.132 & Nayuta (NIC)   \\ 
2019-05-31 & 58634.6 &   16.4  &         --         & 14.954 $\pm$ 0.031 &          --        & Kanata (HONIR) \\
2019-06-03 & 58637.5 &   19.3  & 15.694 $\pm$ 0.156 & 14.795 $\pm$ 0.055 &          --        & Nayuta (NIC)   \\
2019-06-03 & 58637.6 &   19.4  & 16.026 $\pm$ 0.030 & 15.076 $\pm$ 0.495 &          --        & Kanata (HONIR) \\
2019-06-11 & 58645.6 &   27.4  & 15.491 $\pm$ 0.037 &         --         &          --        & Kanata (HONIR) \\
2019-06-13 & 58647.6 &   29.4  & 15.864 $\pm$ 0.056 &         --         &          --        & Kanata (HONIR) \\
2019-06-17 & 58651.6 &   33.4  &         --         &         --         &          --        & Kanata (HONIR) \\
2019-06-24 & 58658.6 &   40.4  &         --         &         --         &          --        & Kanata (HONIR) 
\enddata
\label{tb:nir_log}
\tablenotetext{b}{Relative to the epoch of $B$-band maximum (MJD 58618.24).}
\end{deluxetable*}

Additionally, we downloaded the imaging data obtained by $Swift$ 
Ultraviolet/Optical Telescope (UVOT) from the $Swift$ Data 
Archive\footnote{http://www.swift.ac.uk/swift\_portal/}.
We performed the PSF fitting photometry using {\it IRAF} for these data. 
We calibrated the magnitudes using the comparison stars in the 
$UBV$ bands.
For the $uvw1$ and $uvw2$ bands, we adopted the absolute photometry 
using the zero points reported by \cite{Breeveld11}. 
All $Swift$/UVOT photometric results are listed in Table \ref{tb:uv_log}.

\begin{deluxetable*}{cccccccc} 
\tablewidth{0pt}
\tablecaption{Log of $Swift$/UVOT observations of SN 2019ein.}
\tablehead{
  Date  & MJD  &  Phase\tablenotemark{c}  & $V$   & $B$   & $U$   & $uvw1$ & $uvw2$ \\
        &      &  (day)  & (mag) & (mag) & (mag) & (mag)  & (mag)    
}
\startdata
2019-05-02 & 58605.4 & $-12.8$ & 17.625 $\pm$ 0.128 & 17.594 $\pm$ 0.069 & 17.833 $\pm$ 0.077 & 20.781 $\pm$ 0.448 & 20.667 $\pm$ 0.437 \\
2019-05-03 & 58606.4 & $-11.8$ & 16.570 $\pm$ 0.204 & 16.607 $\pm$ 0.040 & 16.731 $\pm$ 0.063 & 18.591 $\pm$ 0.161 & 19.802 $\pm$ 0.214 \\
2019-05-04 & 58607.8 & $-10.4$ & 15.793 $\pm$ 0.045 & 15.895 $\pm$ 0.037 & 15.788 $\pm$ 0.051 & 17.872 $\pm$ 0.079 & 18.888 $\pm$ 0.109 \\
2019-05-05 & 58608.8 &  $-9.4$ & 15.490 $\pm$ 0.045 & 15.476 $\pm$ 0.034 & 15.317 $\pm$ 0.037 & 17.035 $\pm$ 0.077 & 18.314 $\pm$ 0.066 \\
2019-05-06 & 58609.7 &  $-8.5$ & 15.114 $\pm$ 0.040 & 15.092 $\pm$ 0.035 & 14.956 $\pm$ 0.040 & 16.681 $\pm$ 0.061 & 18.208 $\pm$ 0.074 \\
2019-05-09 & 58612.6 &  $-5.6$ &         --         &          --        &         --         & 15.748 $\pm$ 0.038 & 17.026 $\pm$ 0.054 \\
2019-05-12 & 58615.5 &  $-2.7$ & 14.107 $\pm$ 0.036 & 14.006 $\pm$ 0.026 & 13.912 $\pm$ 0.046 & 15.755 $\pm$ 0.042 & 16.747 $\pm$ 0.048 \\
2019-05-14 & 58617.2 &  $-1.0$ &         --         &          --        &         --         & 16.937 $\pm$ 0.040 & 16.937 $\pm$ 0.040 \\
2019-05-14 & 58617.2 &  $-1.0$ &         --         &          --        &         --         & 16.769 $\pm$ 0.063 & 16.769 $\pm$ 0.063 \\
2019-05-16 & 58619.1 &    0.9  &         --         &          --        &         --         & 15.637 $\pm$ 0.035 & 16.787 $\pm$ 0.055 \\
2019-05-17 & 58620.1 &    1.9  & 13.899 $\pm$ 0.039 & 13.913 $\pm$ 0.038 &         --         & 15.758 $\pm$ 0.052 & 16.755 $\pm$ 0.039 \\
2019-05-26 & 58629.2 &   11.0  &         --         &          --        &         --         &         --         & 18.157 $\pm$ 0.034 \\
2019-05-26 & 58629.3 &   11.1  &         --         &          --        &         --         &         --         & 17.769 $\pm$ 0.075 \\
2019-05-27 & 58630.2 &   12.0  &         --         &          --        &         --         &         --         & 18.078 $\pm$ 0.050  
\enddata
\label{tb:uv_log}
\tablenotetext{c}{Relative to the epoch of $B$-band maximum (MJD 58618.24).}
\end{deluxetable*}

\subsection{Spectroscopy}\label{sec:spec}
We obtained optical spectra of SN 2019ein using HOWPol.
The wavelength coverage is $4500$--$9200$ \AA \, and the wavelength 
resolution is $R = \lambda/\Delta\lambda \simeq 400$ at 6000 \AA. 
For wavelength calibration, we used sky emission lines. 
To remove cosmic ray events, we used the {\it L. A. Cosmic} pipeline 
(\citealt{Dokkum01}; \citealt{Dokkum12}). 
The flux of SN 2019ein was calibrated using the data of 
spectrophotometric standard stars taken on the same nights. 

In addition, we obtained optical spectroscopic data using the Kyoto 
Okayama Optical Low-dispersion Spectrograph with an integral field unit 
(KOOLS-IFU; \citealt{Yoshida05}, \citealt{Matsubayashi19}) installed 
on the 3.8-m Seimei telescope through the optical fibers. 
The Seimei telescope of Kyoto University is a newly established 
telescope at the Okayama Observatory (\citealt{Kurita10}). 
The data were taken under the programs 19A-1-CN07, 19A-1-CN09, 
19A-1-CT01, 19A-K-0003, 19A-K-0004 and 19A-K-0010.
KOOLS-IFU is equipped with four grisms, among which we used the 
VPH-blue.  
The wavelength coverage is $4000$--$8900$ \AA \ and the wavelength 
resolution is $R = \lambda/\Delta\lambda \sim 500$. 
The data reduction was performed using Hydra package in {\it IRAF} 
(\citealt{Barden94}; \citealt{Barden95}) and a reduction software 
specifically developed for KOOLS-IFU data.
For the wavelength calibration, we used arc lamp (Hg and Ne) data.
A lot of the spectroscopic observation is listed in Table 
\ref{tb:spec_log}.

\begin{deluxetable*}{cccccc} 
\tablewidth{0pt}
\tablecaption{Log of the spectroscopic observations of SN 2019ein.}
\tablehead{
  Date  & MJD  &  Phase\tablenotemark{d}  & Coverage  & Resolution  & Telescope  \\
        &      &  (day)  & (\AA)     & (\AA)       & (Instrument)  
}
\startdata
2019-05-03 & 58606.7 & $-11.5$ & $4000$--$8900$  & 500 & Seimei (KOOLS)  \\
2019-05-05 & 58608.8 &  $-9.5$ & $4500$--$9200$  & 400 & Kanata (HOWPol) \\
2019-05-06 & 58609.7 &  $-8.5$ & $4000$--$8900$  & 500 & Seimei (KOOLS)  \\
2019-05-09 & 58612.6 &  $-5.6$ & $4000$--$8900$  & 500 & Seimei (KOOLS)  \\
2019-05-11 & 58614.7 &  $-3.6$ & $4500$--$9200$  & 400 & Kanata (HOWPol) \\
2019-05-12 & 58615.7 &  $-2.6$ & $4000$--$8900$  & 500 & Seimei (KOOLS)  \\
2019-05-14 & 58617.6 &  $-0.6$ & $4000$--$8900$  & 500 & Seimei (KOOLS)  \\
2019-05-21 & 58624.6 &    6.4  & $4500$--$9200$  & 400 & Kanata (HOWPol) \\
2019-05-29 & 58632.5 &   14.2  & $4000$--$8900$  & 500 & Seimei (KOOLS)  \\
2019-06-11 & 58645.6 &   27.4  & $4500$--$9200$  & 400 & Kanata (HOWPol) \\
2019-06-12 & 58646.6 &   28.4  & $4500$--$9200$  & 400 & Kanata (HOWPol) \\
2019-06-19 & 58653.6 &   35.3  & $4500$--$9200$  & 400 & Kanata (HOWPol) \\
2019-06-25 & 58659.6 &   41.4  & $4500$--$9200$  & 400 & Kanata (HOWPol)
\enddata
\label{tb:spec_log}
\tablenotetext{d}{Relative to the epoch of $B$-band maximum (MJD 58618.24). }
\end{deluxetable*}

\section{Results}\label{sec:results}
\subsection{Light Curves}\label{sec:lc}

Based on the spectral similarity to HV SN 2002bo 
in the classification spectrum (\citealt{Burke19}) and further 
analyses of the spectral properties which also suggest the HV SN Ia 
classification for SN 2019ein (see \S\ref{sec:spectra} \& 
\S\ref{sec:hvf2} for details), we especially focus on the comparison 
of its light curves to those of HV SNe Ia.

\begin{figure*}[t]
\centering
\includegraphics[width=16cm,clip]{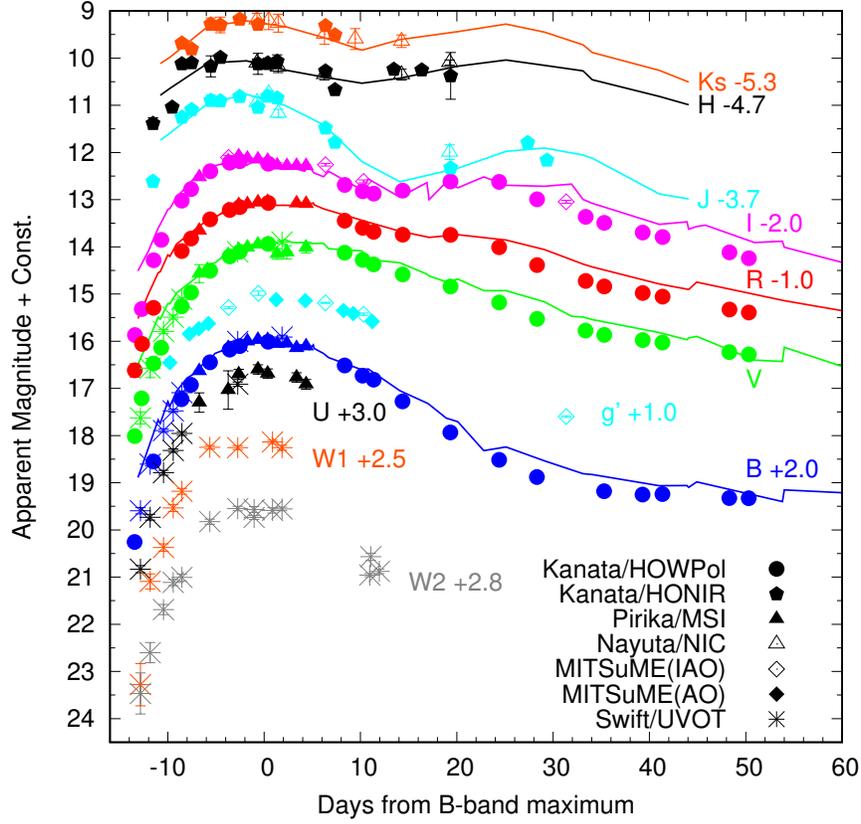}
\caption{Multi-band light curves of SN 2019ein.
The different symbols denote data that were obtained using 
different instruments (see the figure legends).
The light curve of each band is shifted vertically as 
indicated in the figure.
We adopted MJD 58618.24 $\pm$ 0.07 as day zero.
For comparison, we show the light curves of SN 2002bo with 
solid lines (\citealt{Benetti04}; \citealt{Krisciunas04}).}
\label{fig:lc}
\end{figure*}

Figure \ref{fig:lc} shows the multi-band light curves of SN 2019ein. 
We estimate the epoch of the $B$-band maximum as MJD 58618.24 
$\pm$ 0.07 (2019 May 15.2) by performing a polynomial fitting to 
the data points around the maximum light.
In this paper, we refer the $B$-band maximum date as day zero.
We derived the decline rate in the $B$ band as 
$\Delta m _{15}$($B$) = 1.36 $\pm$ 0.02 mag.
As compared to some well-studied HV SNe Ia, SN 2019ein shows a 
faster decline than SNe 2002bo (1.13 $\pm$ 0.05 mag; 
\citealt{Benetti04}), 2002dj (1.08 $\pm$ 0.05 mag; 
\citealt{Pignata08}), and 2004dt (1.21 $\pm$ 0.05 mag; 
\citealt{Altavilla07}).
The decline rate is similar to that of HV SN 2002er 
(1.33 $\pm$ 0.04 mag; \citealt{Pignata04}).
The $B$-band maximum magnitude is 13.99 $\pm$ 0.03 mag.
The maximum date, the peak magnitude, and the decline rate 
in the $BVRI$ bands are listed in Table \ref{tb:para}.

\begin{figure}[t]
\centering
\includegraphics[width=9cm,clip]{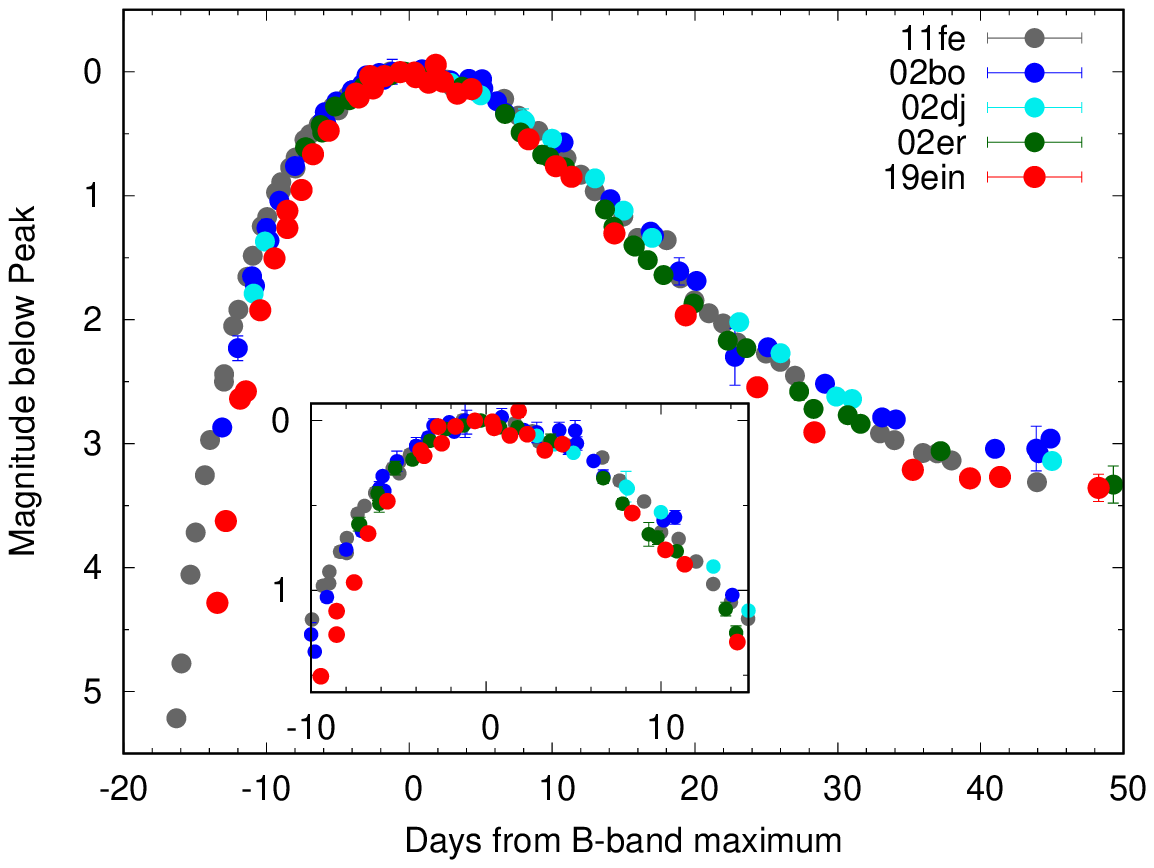}
\caption{$B$-band light curve of SN 2019ein.
For comparison, we plot those of SNe 2002bo (\citealt{Benetti04}; 
\citealt{Krisciunas04}), 2002dj (\citealt{Pignata08}), 2002er 
(\citealt{Pignata04}), and 2011fe (\citealt{Zhang16}).
The inset panel shows the light curves expanded around the 
maximum light.
}
\label{fig:lcb}
\end{figure}

Figure \ref{fig:lcb} shows the $B$-band light curves of 
SN 2019ein, well-studied HV SNe, and NV SN 2011fe.
This highlights that the observation for SN 2019ein has 
started at the most infant phase among HV SNe.
Although the data in the rising part of HV SNe are 
limited, it is clear that SN 2019ein shows a fast rise 
and decline.
The decline of SN 2019ein is as fast as that of SN 2002er.

The decline rate is consistent with other observational properties. 
The time interval between the first peak and the second peak in 
the $I$ band shows a correlation with 
$\Delta m_{15}$($B$) (\citealt{Elias-Rosa08}).
The relation predicts the time interval to be 
24.33 $\pm$ 0.26 days. 
From the light curve, the interval is derived to be 24.76 $\pm$ 
2.82 days, being consistent with that expected from 
$\Delta m_{15}$($B$). 
The line depth ratio of Si {\sc ii} $\lambda$5972 to 
Si {\sc ii} $\lambda$6355 is an indicator of the absolute magnitude 
(\citealt{Nugent95}) or $\Delta m_{15}$($B$) (\citealt{Blondin12}). 
A similar relation also exists for the ratio of the pseudo 
equivalent widths (pEWs). 
These ratios measured for SN 2019ein are also consistent with 
$\Delta m_{15}$($B$) (see \S\ref{sec:spectra}).

\begin{deluxetable}{ccccc}
\tablewidth{0pt}
\tablecaption{Parameters of the $BVRI$-band light curves for SN 2019ein}
\tablehead{
  Band &  Maximum date & Maximum magnitude & $\Delta m_{15}$ & $\Delta$t\tablenotemark{e}\\
  & (MJD) & (mag) & (mag) & (days)
}
\startdata
$I$ & 58616.75 (0.10) & 14.15 (0.02) & 0.76 (0.06) & $-1.49$ \\
$R$ & 58619.18 (0.08) & 14.08 (0.02) & 0.45 (0.16) & +0.94 \\
$V$ & 58619.76 (0.08) & 13.92 (0.02) & 0.76 (0.01) & +1.52 \\
$B$ & 58618.24 (0.07) & 13.99 (0.03) & 1.36 (0.02) & -- 
\enddata
\label{tb:para}
\tablenotetext{e}{The time of difference with $B$-band maximum.}
\end{deluxetable}

\subsection{Absolute Magnitude}\label{sec:absmag}

To estimate the extinction within the host galaxy and the 
absolute magnitude of SN 2019ein, we apply the light curve 
fitter (SNooPy; \citealt{Burns11}) to the light 
curves of SN 2019ein.
We then obtain $E(B-V)_{host} = 0.09 \pm 0.02$ (statistical) $\pm 0.06$ 
(systematic) mag with $R_{V} =$ 1.5 (see below).
As a cross check, we use phenomenological relations to estimate 
the host extinction, based on the $B-V$ color.
The $B-V$ color at the maximum light of SN 2019ein 
is 0.05 $\pm$ 0.02 mag after correcting for the MW extinction with  
$E(B-V)_{MW} = 0.011$ mag and $R_{V} = 3.1$ (\citealt{Schlafly11}). 
We may use the relation between the intrinsic $B-V$ color and the 
decline rate; we obtain $E(B-V)_{host} = 0.09 \pm 0.02$ mag using 
the relation by \citet{Phillips99}, or 
$E(B-V)_{host} = 0.06 \pm 0.09$ mag using that given by \citet{Reindl05}.
We may also use the relation between the line velocity of 
Si {\sc ii} $\lambda$6355 and the intrinsic $B-V$ color at the 
maximum light; we obtain $E(B-V)_{host} = 0.04 \pm 0.07$ mag 
using the relation by \citet{Foley11}, or $0.14 \pm 0.08$ mag 
using that presented by \citet{Blondin12}.
The values estimated by the different methods largely agree with 
each other, and we adopt $E(B-V)_{host} = 0.09 \pm 0.02$ mag obtained 
by SNooPy throughout the paper. 
The estimated extinction is not large, and will not 
affect the main conclusions in the present work.

In Table \ref{tb:distance}, we show the 
$B$-band peak absolute magnitudes of SN 2019ein that are estimated 
using the luminosity-width relations given in the previous papers. 
The derived values for the absolute magnitude agree with each 
other within the uncertainty. 
The average value is $-19.14 \pm 0.10$ mag.
From the estimated peak absolute magnitude,  
$E(B-V)_{host} = 0.09 \pm 0.02$ mag and $R_{V} = 1.55 \pm 0.06$ 
(as is frequently adopted for HV SNe; \citep{Wang09}), 
we obtain the distance to SN 2019ein (NGC5353) as 
$\mu = 32.95 \pm 0.12$ mag. 
The value is larger than that from the NASA/IPAC Extragalactic Database 
(NED)\footnote{http://ned.ipac.caltech.edu/} ($\mu = 32.20 \pm 0.54$ mag). 
However, the distance measurements for NGC 5353 found in NED are highly 
uncertain\footnote{The mean value of the results from 
different methods is $\mu = 32.20 \pm 0.54$ mag. A typical error associated 
with each measurement also exceeds $\sim 0.4$ mag.}, and we think that 
the distance obtained here assuming that SN 2019ein follows the SN Ia 
luminosity-width/color relation is indeed more robust than the previous 
estimates; SN 2019ein shows various properties in the light curves and 
the spectra that fit into the correlations found for a sample of SNe Ia, 
and therefore it is rational to assume that the SN also follows the 
standardized candle relation. 
In what follows, we use $E(B-V)_{host} = 0.09 \pm 0.02$ mag 
and $\mu = 32.95 \pm 0.12$ mag unless otherwise mentioned.

\subsection{Spectral Properties}\label{sec:spectra}

\begin{figure*}[t]
\centering
\includegraphics[width=16cm,clip]{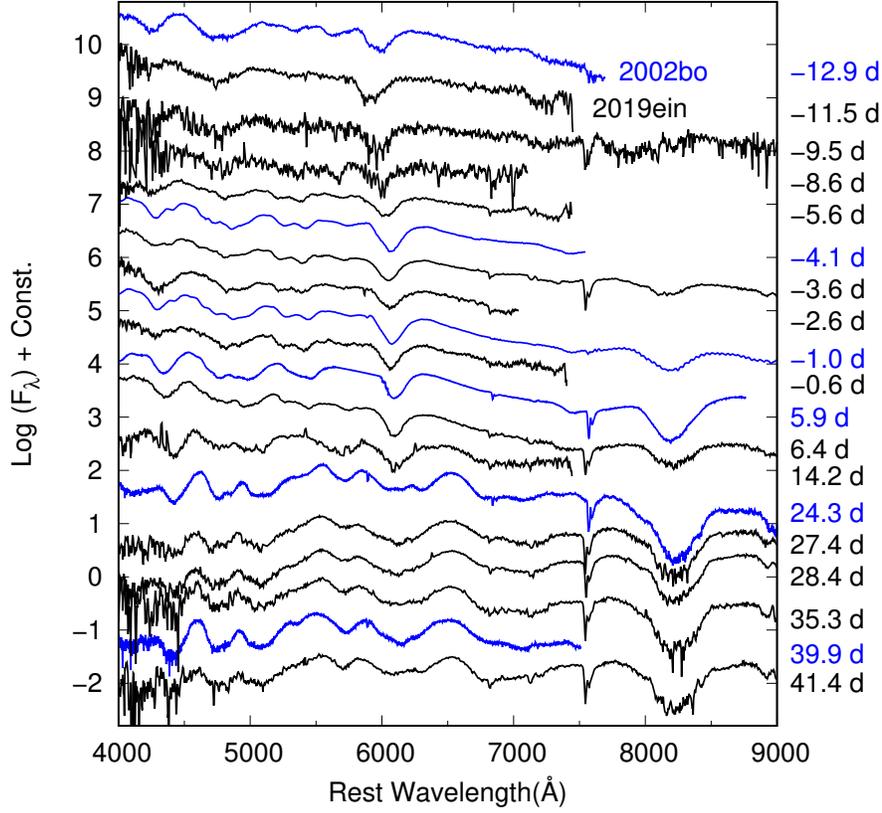}
\caption{Spectral evolution of SN 2019ein (black lines).
The epoch for each spectrum is indicated on the right outside 
the panel.
For comparison, we plot the spectra of SNe 2002bo (blue lines; 
\citealt{Benetti04}, \citealt{Blondin12}).
}
\label{fig:spec}
\end{figure*}

Figure \ref{fig:spec} shows the optical spectra of SN 2019ein 
from $-11.5$ days through $41.4$ days.
The spectra are characterized by absorption lines of Si {\sc ii}, 
S {\sc ii}, Fe {\sc ii}, Fe {\sc iii} and the Ca {\sc ii} IR triplet.
The spectra of SN 2019ein show a striking similarity to those of 
HV SN 2002bo. 

\begin{figure}[t]
\centering
\includegraphics[width=9cm,clip]{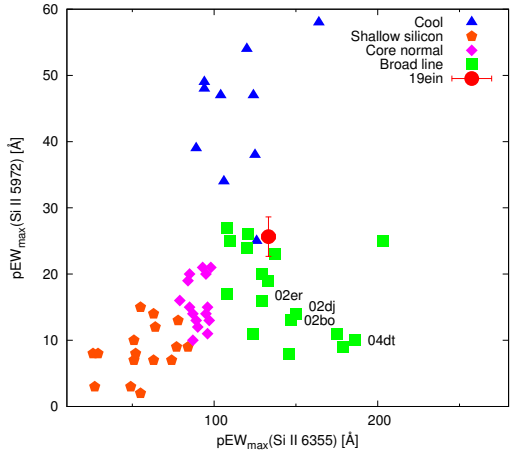}
\caption{Pseudo equivalent widths of 
Si {\sc ii} $\lambda$6355 and Si {\sc ii} $\lambda$5792.
The comparison samples are from \cite{Silverman12a}.
Note that the Broad Line and Core Normal classes correspond 
to the HV and NV classes.
}
\label{fig:pEW}
\end{figure}

\cite{Branch06} suggested that SNe Ia are 
classified into four subgroups using the EWs of Si {\sc ii} 
$\lambda$5792 and Si {\sc ii} $\lambda$6355. 
The pEWs of Si {\sc ii} $\lambda$5792 and Si {\sc ii} $\lambda$6355 
measured for SN 2019ein at $-0.6$ days are 133.24 $\pm$ 4.16 \AA \ 
and 25.65 $\pm$ 2.98 \AA, respectively. 
In Figure \ref{fig:pEW}, we show the \citet{Branch06} diagram.
In this classification scheme, SN 2019ein is 
classified as the ``Broad Line (BL)'' subgroup, the distribution 
of which overlaps with HV SNe (\citealt{Wang09}).
The pEW ratio of Si {\sc ii} $\lambda$5972 and 
Si {\sc ii} $\lambda$6355 and the depth ratio of these lines 
are 0.23 $\pm$ 0.03 and 0.19 $\pm$ 0.02, respectively.
From Figure 14 of \citet{Blondin12}, these ratios are consistent 
with $\Delta m_{15}$($B$).
SN 2019ein is thus classified as a HV (and BL) 
SN Ia from its maximum-light spectrum, irrespective of the nature 
of the spectra in the pre-maximum phase.

\begin{figure}[t]
\centering
\includegraphics[width=9cm,clip]{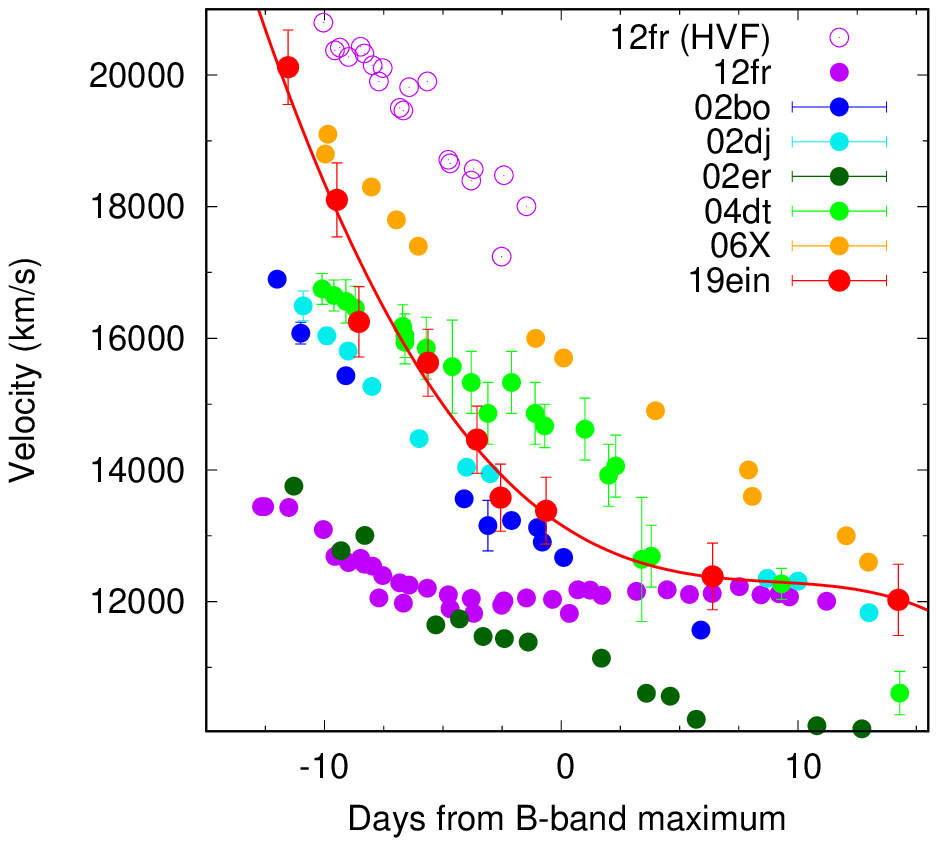}
\caption{Evolution of line velocity of Si {\sc ii} $\lambda$6355 
in SN 2019ein.
For comparison, we plot those in SNe 2002bo (\citealt{Benetti04}; 
\citealt{Blondin12}; \citealt{Silverman12a}), 
2002dj (\citealt{Pignata08}), 2002er (\citealt{Kotak05}; 
\citealt{Silverman12a}; \citealt{Blondin12}), 2004dt 
(\citealt{Altavilla07}), 2006X (\citealt{Yamanaka09}), and 2012fr 
(\citealt{Childress13}; the open and filled circles denote the 
velocity of the high velocity feature component and the photospheric 
component, respectively).
The red solid line is a polynomial function that fits to the data 
of SN 2019ein.
}
\label{fig:sivel2}
\end{figure}

We therefore apply a single-Gaussian fit to the absorption 
line profiles to study the evolution of the line velocities, in the 
same manner applied to HV SNe in the previous studies\footnote{A 
possible contamination by HVFs is discussed in \S\ref{sec:hvf2}}.
For each spectrum, the fit is performed several times with the 
continuum determination varied, using the splot task in {\it IRAF}.
We determined the uncertainty from the root mean square of the 
standard deviation in the fits and the wavelength resolution.
In Figure \ref{fig:sivel2}, we compare the evolution of the line 
velocity of Si {\sc ii} $\lambda$6355 to those of other HV 
SNe, and to that of SN 2012fr which showed strong HVFs. 
SN 2019ein is among SNe which show the fastest line velocities in 
the early phase. 
The line velocity of SN 2019ein is $\sim$ 20,000 km s$^{-1}$ at 
$-12$ days, and it is as fast as those of SN 2006X and the HVF 
component seen in SN 2012fr.
SN 2019ein shows the rapid evolution in the velocity 
toward the $B$-band maximum. 
The velocity of SN 2019ein is decreased to $\sim$ 14,000 km s$^{-1}$ 
at $-0.6$ days. 
Around the maximum light, it becomes similar to that of SN 2002bo, 
despite initially the larger velocity. 
After the maximum light, the velocity evolution of SN 2019ein is 
flattened at $\sim 12,000$ km s$^{-1}$. 

Figure \ref{fig:sivel} shows an expanded view for the Si {\sc ii} 
$\lambda$6355 feature at several phases.
Initially at $\sim -12$ days, the feature resembles that of 
SN 2012fr, which is dominated by the HVF component. 
However, subsequent evolution is different; SN 2012fr developed 
two distinct features interpreted as the detached  
HVF and the photospheric components, where the latter shows 
slow evolution in its velocity (\citealt{Childress13}). 
On the other hand, SN 2019ein does not develop clear 
multiple features, and a single component seems 
to continuously move toward the lower velocity as time goes by
(see \S\ref{sec:hvf2} for further details). 
This property is consistent with the evolution of the Si {\sc ii} 
in HV SNe. 
In the first epoch, the velocities of the Si {\sc ii} seen in 
HV SNe 2002bo, 2002dj, and 2004dt are slower than that of SN 2019ein.
However, the line width and strength are almost the same in 
these SNe Ia including SN 2019ein, again confirming the 
classification of SN 2019ein as a HV SN Ia. 
At $-9$ days, SN 2019ein rather resembles SNe 2002bo and 2002dj, 
and is clearly different from SN 2012fr. 
The photospheric component of SN 2012fr is now as noticeable as 
the HVF one, but such a behavior is never seen for SN 2019ein; 
while the spectrum of SN 2019ein at $-9.5$ days is noisy, it is 
still consistent with a single component. 
Toward the maximum light, SN 2019ein becomes more and more similar 
to SNe 2002bo and 2002dj, and the difference between SNe 2019ein 
and 2012fr becomes more noticeable. 
Among the sample, SN 2004dt shows a broad and boxy Si {\sc ii} 
line profile unlike the other comparison SNe Ia. 
We note that SN 2004dt is indeed suggested to be an outlier 
in the HV class (\citealt{Maeda10a}). 

We estimate the velocity gradient of Si {\sc ii} $\lambda$6355.
In this paper, we obtain the velocity gradient using the method of 
\citet{Foley11}, in which it is estimated by fitting a liner function 
to the velocity evolution within the fixed time interval 
($-6$ days $\leq$ t $\leq$ 10 days).
Following this definition, we derive the velocity gradient of 
SN 2019ein as 235.3 $\pm$ 61.7 km s$^{-1}$ day$^{-1}$.
By the same method, we also estimate the velocity gradient of 
other SNe Ia 
as follows; SNe 2002bo ($215.1 \pm 23.3$ km s$^{-1}$ day$^{-1}$), 
2004dt ($243.5 \pm 18.2$ km s$^{-1}$ day$^{-1}$), 
2006X ($239.6 \pm 18.4$ km s$^{-1}$ day$^{-1}$), 
2002er ($119.7 \pm 14.5$ km s$^{-1}$ day$^{-1}$), and 
2002dj ($136.4 \pm 15.1$ km s$^{-1}$ day$^{-1}$). 
SN 2019ein shows the velocity gradient similar to SNe 2002bo, 2004dt, 
and 2006X, and larger than SNe 2002er and 2002dj.

The velocity gradient defined in the maximum light has 
been suggested to be correlated to the Si velocity at the 
maximum light (\citealt{Foley12a}; \citealt{Wang13}).
SN 2019ein fits into this relation. 
However, the present study shows that it is not necessarily 
the case for the pre-maximum, sufficiently early phases. 
Despite the lower Si velocity at the maximum light than SNe 2006X 
and 2004dt, the velocity of SN 2019ein evolves much faster in the 
pre-maximum phase than these SNe. 
Because of this evolution effect, the Si {\sc ii} $\lambda$6355 velocity 
in the early phase of SN 2019ein is among the largest in the HV SN 
group, despite only a moderately large velocity at the maximum light.
This indicates that the diversity exists within the HV SN group.
While the speed of the spectral evolution around the 
maximum light is highly dependent on the velocity observed there, it is 
more strongly correlated with $\Delta m_{15}(B)$ (or generally the speed 
of the evolution in the light curves) in the pre-maximum phase.
We note that such a correlation between the spectral evolution and 
$\Delta m_{15}(B)$ is indeed well known for NV SNe (e.g., 
\citealt{Benetti05}), but has not been clarified for HV SNe due to the 
limited sample.

\subsection{High Velocity Features?}\label{sec:hvf2}

\begin{figure*}[t]
\centering
\includegraphics[width=13cm,clip]{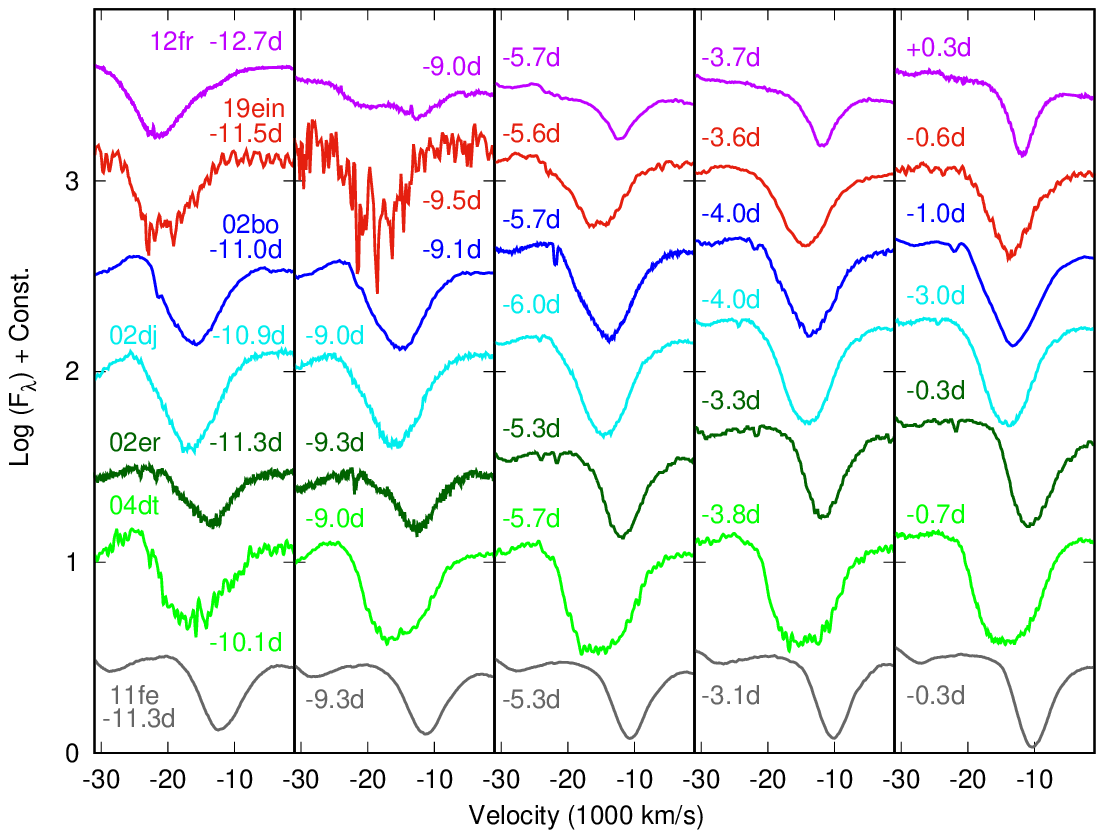}
\caption{Evolution of Si {\sc ii} $\lambda$6355 
line profile of SN 2019ein as shown in velocity space.
For comparison, we show the spectra of SNe 2012fr 
(\citealt{Childress13}), 2002bo (\citealt{Benetti04}; 
\citealt{Blondin12}), 2002er (\citealt{Kotak05}), 
2002dj (\citealt{Pignata08}), 
2004dt (\citealt{Altavilla07}), and 2011fe (\citealt{Pereira13}).}
\label{fig:sivel}
\end{figure*}

\begin{figure}[t]
\centering
\includegraphics[width=9cm,clip]{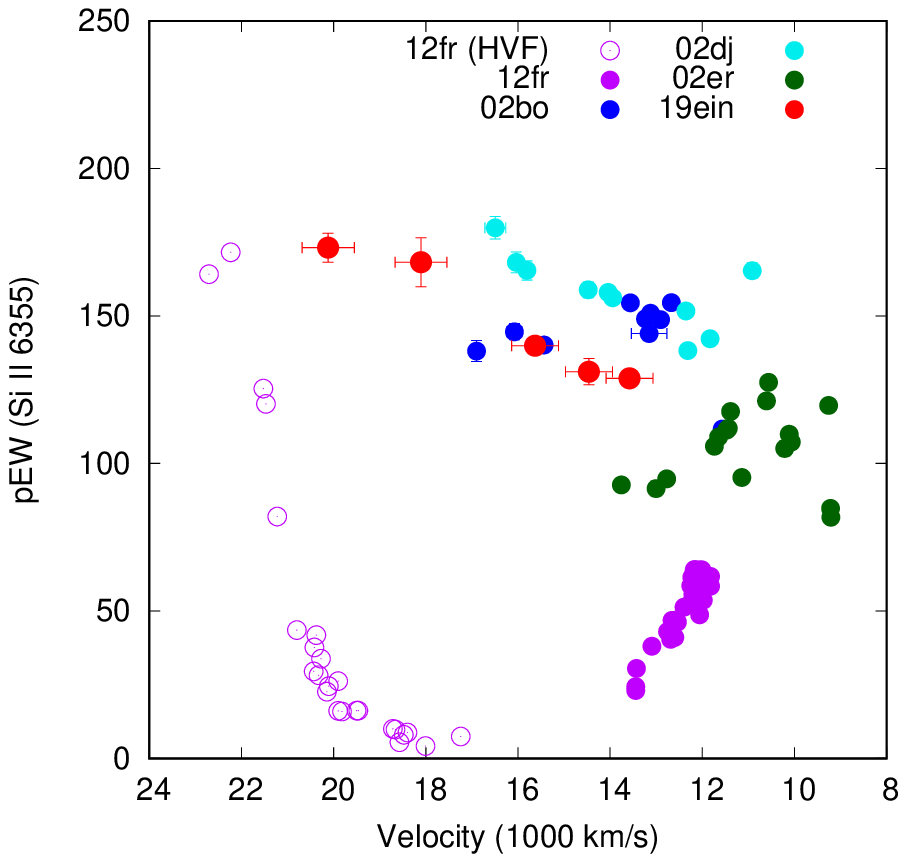}
\caption{Evolution of pseudo equivalent width v.s. velocity of 
Si {\sc ii} $\lambda$6355 for SN 2019ein.
For comparison, we plot those of SNe 2002bo (\citealt{Benetti04}; 
\citealt{Blondin12}; \citealt{Silverman12a}), 2002dj 
(\citealt{Pignata08}),  
2002er (\citealt{Kotak05}; \citealt{Silverman12a}), and  
2012fr (\citealt{Childress13}; the open and filled circles denote 
the velocity of the 
high velocity feature component and the photospheric component, respectively).}
\label{fig:velpew}
\end{figure}

It has been suggested that most, if not all, 
of SNe Ia show the so-called HVF detached from the photospheric 
component, if spectra are taken sufficiently early.
For NV SNe, the HVF and photospheric  
components can be well separated in velocity space (e.g., see 
Figure \ref{fig:sivel} for SN 2012fr). 
This may not be the case for HV SNe where the line blending between 
two components, even if the HVF exists, can be significant. 
As such, one may ask if the high-velocity `photospheric' component in 
the earliest phases may be contaminated substantially by the possible 
HVF, which would overestimate the absorption velocity derived from a 
single-component fit. 
In this section, we further investigate a possibility if the HVFs could 
contaminate to the early phase spectra, and if this contamination would 
mimic the high-velocity `single component'. 
Note that this is indeed a generic issue for HV SNe, not only for 
SN 2019ein.
In Figure \ref{fig:velpew}, we show the evolution 
of the pEW and velocity of Si {\sc ii} $\lambda$6355.
In the case of SN 2019ein, the pEW gradually decreases as time 
goes by. 
SNe 2002bo and 2002er show similar evolution.
The behavior is very different from that found for SN 2012fr.

\begin{figure}[t]
\centering
\includegraphics[width=9cm,clip]{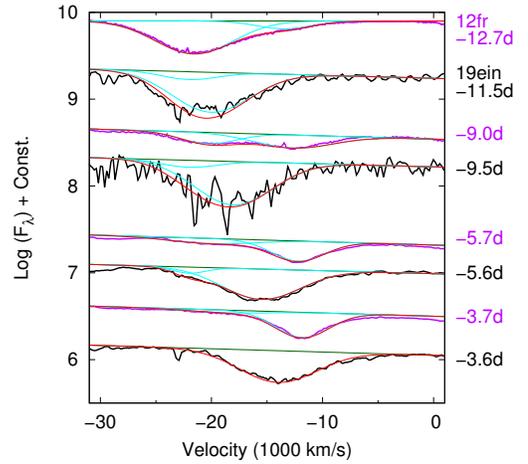}
\caption{Two-component fit to 
Si {\sc ii} $\lambda$6355 line of SN 2019ein (black lines) and 
to that of SN 2012fr (\citealt{Childress13}; purple lines) before the 
maximum light.
We plot the pseudo-continuum and two individual components as  
the dark green and cyan lines, respectively.
The sums of these components are shown as the red lines.}
\label{fig:hvf}
\end{figure}

To further quantify the issue, we fit the profile of 
Si {\sc ii} $\lambda$6355 with a combination of two-Gaussian functions, 
for SNe 2012fr and 2019ein.
The early phase spectra of SN 2012fr show obviously distinct two components.
It is seen that the FWHMs of the HVF and photospheric components both 
remain almost unchanged over time (\citealt{Childress13}).
At the maximum light, the Si {\sc ii} $\lambda$6355 line shows the 
photospheric component only.
We thus measure the FWHM of the photospheric component at this phase, and 
use the same FWHM for the other epochs.
With the FWHM of the photospheric component fixed, we vary the remaining 
parameters (the velocities of both components, the FWHM of the HVF, and 
the relative depth) to fit each spectrum.
Figure \ref{fig:hvf} shows that the spectra of SN 2012fr before the 
maximum light are well fitted by the combination of the HVF and 
photospheric components.
The results of the fits (e.g., the velocities of the two components) are 
consistent with those of \citet{Childress13}.

The two-Gaussian fit to the spectra of SN 2019ein follows 
the same manner.
First, we fit the FWHM of the photospheric component using the maximum 
light spectrum.
With this fixed, the velocity of the photospheric component can be well 
constrained for each spectrum, by fitting the line profile in the red side 
where the contribution by the possible HVF should be essentially zero.
The residual in the blue side is then practically 
associated to the HVF, which allows to constrain the velocity and the 
FWHM of the possible HVF at each epoch.
Since the spectra of SN 2019ein are well fitted by a single-Gaussian 
function, obviously the two-component fit works well as (Figure 
\ref{fig:hvf}).
However, the contribution by the HVF is negligibly small at all the epochs.
The velocity and the pEW of Si {\sc ii} $\lambda$6355 (in the photospheric  
component) are little influenced by introducing the additional HVF 
contribution.

This result is consistent with the previous statistical 
study of the HVFs, given small $\Delta m_{15}$($B$) for SN 2019ein; 
the HVFs are less frequently found, at a given epoch, for more 
rapidly declining SNe (\citealt{Childress13}; \citealt{Zhao16}).
Detecting the HVFs for the (relatively) rapid 
decliner would require extremely rapid follow-up spectroscopy. 
Indeed, the first spectrum for SN 2019ein reported by the LCO Global 
SN project, taken within a day of the 
discovery\footnote{https://wis-tns.weizmann.ac.il/object/2019ein} 
(\citealt{Burke19}), 
shows clear HVFs both for the Ca {\sc ii} H\&K and Ca {\sc ii} NIR, 
the latter perhaps contaminating the HVF of O {\sc i} $\lambda$7774. 
There is further a hint of the HVF of Si {\sc ii} $\lambda$6355, 
which shows a faster velocity than Si {\sc ii} $\lambda$5972. 
The characteristic velocity of these HVFs is 
$\sim 25,000 - 30,000$ km s$^{-1}$.
The velocities seen in these clear HVFs are even much 
faster than the velocities of the photospheric component studied here, 
by $\sim 10,000$ km s$^{-1}$, and thus it is nearly impossible that it 
would substantially  affect our analysis.

In summary, SN 2019ein belongs to the HV SN Ia group. 
The very high velocity found for SN 2019ein 
probably stems from the very early discovery and follow-up.
SN 2019ein therefore serves as the best 
studied example of a HV SN with relatively large $\Delta m_{15}(B)$. 

\section{Discussion}\label{sec:disc}

\begin{deluxetable}{cc}
\tablewidth{0pt}
\tablecaption{Estimated peak absolute magnitudes in the 
$B$ band for SN 2019ein.}
\tablehead{$M_{B}$ (mag) & Reference}
\startdata
 $-19.10 \pm 0.23$ & \citet{Phillips99}\\
 $-19.18 \pm 0.24$ & \citet{Altavilla04}\\
 $-19.12 \pm 0.24$ & \citet{Wang05}\\
 $-19.15 \pm 0.14$ & \citet{Prieto06} 
\enddata
\label{tb:distance}
\end{deluxetable}
 

\subsection{Light Curves in the Rising Phase}\label{sec:rise_time}

We obtained the multi-band light curves of SN 2019ein from the 
early rising part presumably soon after the explosion. 
From these data, we try to constrain the explosion date and the 
rise time.
We assume the homologously expanding ``fireball model'' 
(\citealt{Arnett82}; \citealt{Riess99}; \citealt{Nugent11}) to 
estimate the explosion date. 
In this model, the luminosity/flux ($f$) increases as 
$f \propto t^{2}$, where $t$ is the time since the zero point. 
In this paper, we assume that the zero point in the time axis in 
this relation is the same for different bands (i.e., the explosion 
date), and adopt the same power-law index of 2 in all the 
$BVRI$ bands. 

\begin{figure}[t]
\centering
\includegraphics[width=9cm,clip]{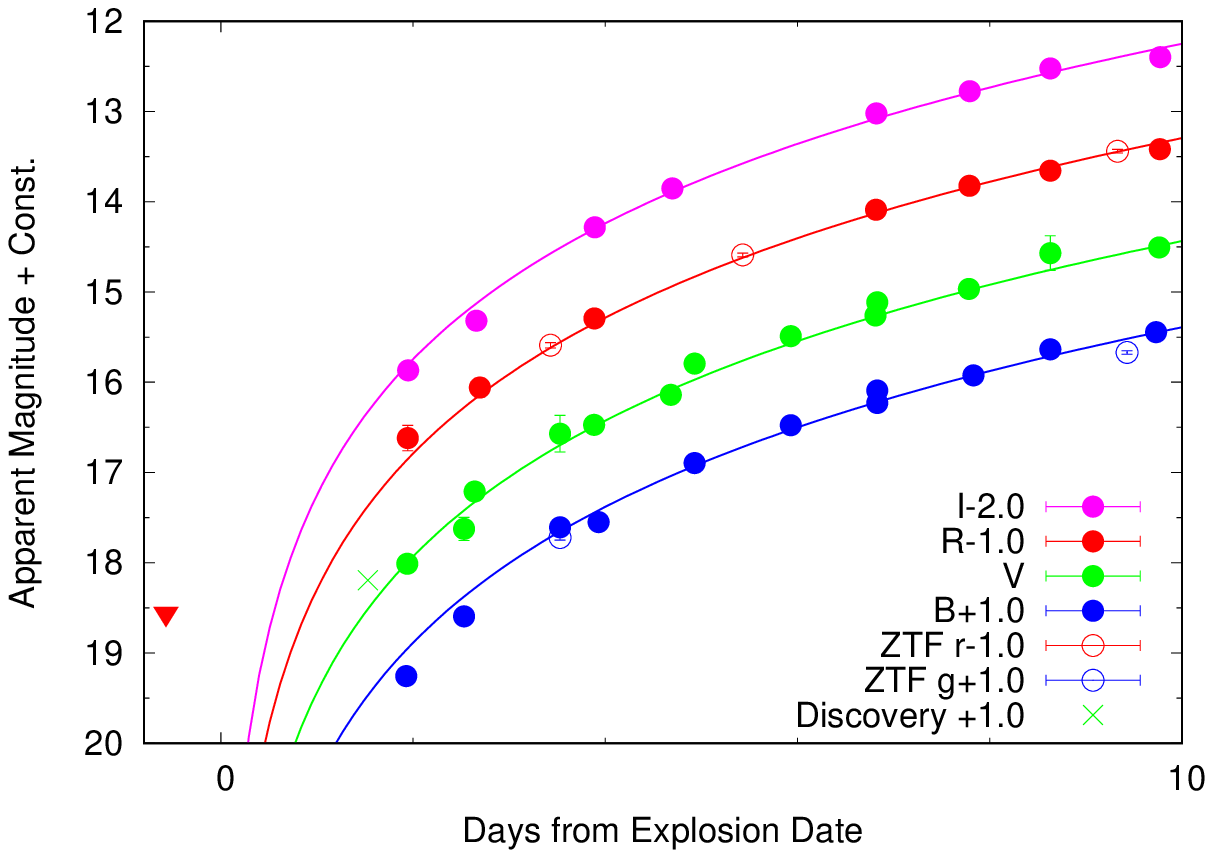}
\caption{Optical light curves of SN 2019ein in the early phase.
We also plot the discovery magnitude (the cross symbol, 
\citealt{Tonry19}) and the data obtained by ZTF taken from the 
Transient Name Server.
The red and blue open circles denote data in the $r$ and $g$ bands, respectively. 
The upside-down triangle indicates the upper-limit magnitude 
obtained by ZTF.
The explosion date is estimated to be MJD 58602.87 $\pm$ 0.55
using the quadratic function (see \S\ref{sec:rise_time}).}
\label{fig:comp_rising}
\end{figure}

In Figure \ref{fig:comp_rising}, we show that the early light 
curves are well fitted by the fireball model. 
The rising part of SN 2019ein, in all the $BVRI$ bands, 
can be explained by this simple fireball model. 
By the fitting, we estimate the explosion date as MJD 58602.87 
$\pm$ 0.55.
There is no significant deviation from the fireball model, at 
least after $\sim 2$ days from the putative explosion date. 
No excess is found with respect to the fireball prediction. 
There is either no need to introduce a broken power law in the 
early phase. 

From the earliest portion of the rising light curve, we now 
discuss implications for the progenitor system. 
The ejecta of SNe Ia are expected to collide with a companion 
star, and create additional heat and thermal energy.
The thermal energy is then lost quickly due to the adiabatic 
expansion (e.g., \citealt{Kasen10}), but a small fraction will be 
leaving the system as radiation. 
The expected strength and duration of this emission are larger for 
a more extended companion star. 
Therefore, by the non-detection of such an emission signal, one can 
place an upper limit on the radius of the companion star. 
We constrain the radius of the  companion star of SN 2019ein by 
scaling the parameters found in the previous analysis conducted for 
SN 2011fe (\citealt{Nugent11}).
They estimated the luminosity of SN 2011fe in the early phase from  
$g$-band observation.
At 1.9 days from the explosion, the luminosity of SN 2019ein 
is (2.7 -- 3.6) $\times$ 10$^{41}$ erg s$^{-1}$. 
By scaling the result of \citet{Nugent11} using the relation 
suggested by \citet{Kasen10}, we obtain the upper limit of 
the companion's radius as 4.3 -- 7.6 $R_{\odot}$. 
It thus excludes a red giant companion oriented along 
the line of sight.
We however note that this does not rule out an 
evolutionary scenario where a giant companion star has already 
evolved to a WD (\citealt{Justham11}; \citealt{DiStefano12}; 
\citealt{Hachisu12}).

The rise time, which is a measure of the time interval between the 
explosion and the maximum light, may also provide some insights into 
the explosion property, and its correlation to other observational 
features have been investigated (e.g., \citealt{Hayden10}; 
\citealt{Ganeshalingam11}; \citealt{Jiang20}). 
The rise time, defined as the $B$-band maximum date minus the 
estimated explosion date, is 15.37 $\pm$ 0.55 days for SN 2019ein. 
It is shorter than that of SN 2002bo (17.9 $\pm$ 0.5 days; 
\citealt{Benetti04}) or SN 2002er (18.7 days; \citealt{Pignata04}).

There is a tendency that SNe Ia that have smaller 
$\Delta m_{15}$($B$) (i.e., slower decline) show the longer rise 
time (e.g., \citealt{Perlmutter98}; 
\citealt{Hayden10}; \citealt{Ganeshalingam11}).
The rise time derived here for SN 2019ein indeed 
fits to the distribution of $\Delta m_{15}$ vs. the rise time for 
the HV SNe in Figure 6 of \citet{Ganeshalingam11}.
SN 2002bo has smaller $\Delta m_{15}$($B$) than SN 2019ein, 
and it is consistent with SN 2002bo having the longer rise time. 
Indeed, in this respect, SN 2002er, with similar 
$\Delta m_{15}$($B$) to SN 2019ein but longer rise time than 
SN 2002bo, does not recover the correlation between 
$\Delta m_{15}(B)$ and the rise time (\citealt{Pignata04}). 

This might indicate that $\Delta m_{15}(B)$ is not a single 
function that determines the rise time. 
For example, the velocity may have a role as a second parameter, given 
the diversity in the velocities seen for HV SNe; it is faster for 
SN 2019ein than SN 2002er (Figure \ref{fig:sivel2}) despite similar 
$\Delta m_{15}(B)$.
Indeed, HV SNe appear to have the faster rise time in 
the $B$ band than NV SNe (\citealt{Pignata08}; \citealt{Zhang10}; 
\citealt{Ganeshalingam11}) for given $\Delta m_{15}(B)$.
In summary, we find that the evolution of HV SNe in 
the pre-maximum phase, both spectroscopically and photometrically, is 
controlled not only by $\Delta m_{15}(B)$ but also by the velocity 
(at the maximum light). 
The evolution of the light curve and the velocity may 
provide an important constraint on the nature of the ejecta structure.
Importantly, it does not form a one parameter family. 

\subsection{Structure of the Outermost Layers}\label{sec:km1}

A spectral time series of SNe provides an opportunity to study 
the composition structure within the ejecta (\citealt{Stehle05};  
\citealt{Mazzali08}; \citealt{Tanaka11}; \citealt{Sasdelli14}). 
Given that the photosphere generally recedes in velocity space 
as time goes by due to the density decrease, one can probe the 
property of the outer layer from the spectra taken earlier. 
The spectra presented here start with $\sim -12$ days and then 
$\sim - 10$ days, which are comparable to the phase of the first 
spectrum taken for the well-studied HV SN Ia 2002bo. 
Indeed, given the rapidly evolving nature of SN 2019ein 
(\S\ref{sec:lc}) and the short rise time (\S\ref{sec:rise_time}), 
the estimated explosion date suggests that these spectra were 
obtained shortly after the explosion; $\sim 3.7$ and $\sim 5.7$ days 
after the explosion for these spectra. 
These are among the most infant spectra taken for SNe Ia, 
especially as a HV SN. 
We note that the first classification-report spectrum was taken 
already at 0.82 days since the discovery, which is likely 
$\sim 2.4$ days since the explosion (\S \ref{sec:hvf2}). 

We use two of our spectra taken at the earliest phases to study 
the composition structure of the outermost layer of SN 2019ein, 
which may hint on the general property of HV SNe.
We note that the possible decomposition of the 
absorption lines into the photosperic and HVF components 
(\S \ref{sec:hvf2}) is merely an `interpretation', and the analysis 
here is totally independent from such an issue.
In doing this, we perform one-dimensional (spherically symmetric) 
spectral synthesis calculations using 
TARDIS\footnote{https://tardis-sn.github.io/tardis/} 
(\citealt{Kerzendorf14}).
To simplify the problem, we adopt a power-law density structure 
above 17,000 km s$^{-1}$, with the normalization 
matched to the classical W7 model at this inner boundary. 
While there is a drop of the density in the W7 model beyond 
$\sim 20,000$ km s$^{-1}$, we keep the same density slope 
beyond this. 
We note that there is a diversity seen in this outermost density 
structure for different models, and our density structure is well 
within the model predictions (see, e.g., Figure 3 of 
\citealt{Mazzali14}).
Given the observed spectral lines showing high velocity in the 
earliest spectra of SN 2019ein, at least a moderate amount of 
material above 20,000 km s$^{-1}$ is required.

Given the large degree of freedom even in the 1D model and the 
non-linear nature of the problem to solve, there is no guarantee 
that the spectral synthesis model provides a unique solution.
After testing various possibilities, we have decided to adopt the 
simplest approach, based on underlying nuclear-burning physics. 
In our model, we consider only two characteristic layers; an inner 
O-Ne-C burning layer and an outer unburnt C+O layer. 
In the O-Ne-C burning layer, the mass fractions are set as 
follows; 0.68 (O), 0.025 (Ne), 0.1 (Mg), 0.2 (Si), 0.025 (S), 
and 5e-4 (Ca). 
In the unburnt layer, we adopt the following; 0.5 (C), 0.475 (O), 
and 0.025 (Ne). The solar abundances are added for the heavier 
elements in both layers. 
Note that these compositions well represent those in the outermost 
layers found in the hydrodynamic and nucleosynthesis simulations, 
without introducing mixing between different layers. 
In the W7 model, these two layers are found at 
$\sim 13,000 - 15,000$ km s$^{-1}$ and $> 15,000$ km s$^{-1}$ 
for the O-Ne-C burning and unburnt regions, respectively. 
In the delayed-detonation model CS15DD2 of \citet{Iwamoto99}, 
these are $\sim 18,000 - 30,000$ km s$^{-1}$ and 
$> 30,000$ km s$^{-1}$, respectively. 

\begin{figure}[t]
\centering
\includegraphics[bb=10 90 200 250,clip]{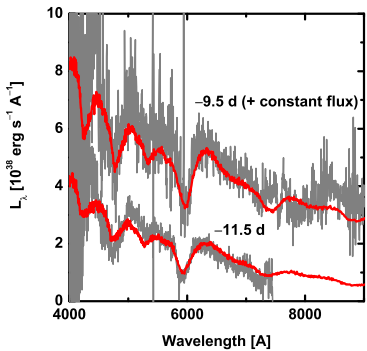}
\caption{Comparison of the observed spectra at $-11.5$ and 
$-9.5$ days (gray lines) to the synthesized spectra (red lines) 
calculated with TARDIS.
These observed spectra are corrected for the MW and 
the host galaxy extinction.
}
\label{fig:km1}
\end{figure}

Figure \ref{fig:km1} shows how this simple model can indeed reproduce 
the earliest-phase spectra of SN 2019ein reasonably well. 
Note that we do not aim at obtaining a detailed fit, 
since we restrict ourselves to the simple model without fine-tuning.
We assume that the explosion date is 1.6 days before the 
discovery, the distance modulus is 33.0 mag, and $E(B-V) = 0.1$ mag and 
$R_V = 1.5$. 
We determine the photospheric velocity to be $20,000$ km s$^{-1}$ 
and $17,000$ km s$^{-1}$ at $-11.5$ and $-9.5$ days (3.7 and 5.7 
days since the assumed explosion date), respectively. 
The photospheric velocities found here are extremely high; these 
are $\sim 13,000 - 14,000$ km s$^{-1}$ and $\sim 12,000 - 13,000$ 
km s$^{-1}$ at similar phases (since the explosion) found for the 
best-studied NV SN Ia 2011fe (\citealt{Mazzali14}), confirming 
that HV SNe show exceedingly higher velocity in the earlier phases. 
The velocities found here is even higher than those found for the 
prototypical HV SN 2002bo at similar epochs, as already indicated 
by the spectral comparison. 
This indicates a large diversity among the HV SN class, and places 
SN 2019ein into the most extreme example in terms of the line 
velocity in the pre-maximum phase. 
This is likely related to the rapidly evolving nature of SN 2019ein 
in its light curves (\S\ref{sec:lc} and \S\ref{sec:rise_time}).

\begin{figure*}
\begin{center}
 \begin{minipage}{0.4\hsize}
   \includegraphics[bb=10 90 200 250,clip]{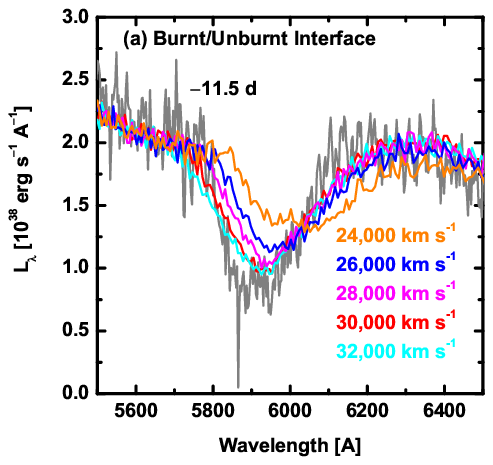}
 \end{minipage}
 \begin{minipage}{0.4\hsize}
   \includegraphics[bb=10 90 200 250,clip]{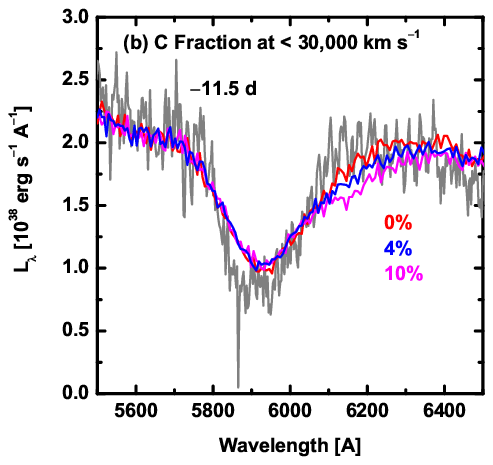}
 \end{minipage}
\end{center}
\caption{Comparison between the observed Si {\sc ii} $\lambda$6355 
at $-11.5$ days (gray line) and the synthesized spectra.
The observed spectrum is corrected for the MW and the host 
galaxy extinction.
(a) Shown here are models with different velocities at the 
interface between the burnt and unburnt layers. 
The dividing velocities are denoted in the figure legends.
(b) Shown here are models with different mass fractions of carbon 
in the O-Ne-C burning layer.
Three lines denote the 0\%, 4\% and 10\% for the carbon fraction, 
respectively.
}
\label{fig:km2}
\end{figure*}

Figure \ref{fig:km2} shows how the velocity at the interface 
between the burnt and unburnt layers can be constrained. 
We note that the overall spectral appearance is not sensitive to 
this velocity, but the line profile does. 
As long as this velocity is larger than $\sim 30,000$ km s$^{-1}$, 
the Si {\sc ii} profile does not change significantly. 
On the other hand, once the interface velocity is set below 
$\sim 30,000$ km s$^{-1}$, the line profile shifts to the lower 
velocity as this interface is moved toward a deeper region. 
From this exercise, we place the constraint that the burnt/unburnt 
interface is placed at $> 30,000$ km s$^{-1}$. Note that we do not 
require, indeed do disfavor, the mixing of the inner, more advanced 
burning region (such as the Si burning) into the region above  
$\sim 17,000$ km s$^{-1}$ as we probe using these spectra. 

To further constrain the amount of unburnt carbon below 
30,000 km s$^{-1}$, we also vary the mass fraction of carbon 
in the O-Ne-C burning layer ($17,000 - 30,000$ km s$^{-1}$). 
The default value, 0\% of carbon contamination, 
works well to reproduce the line profile in the red side. 
Note that while the carbon fraction is set constant within the 
O-Ne-C burning region, the constraint is basically placed at the 
region close to the photosphere, i.e., $\sim 20,000$ km s$^{-1}$. 
By increasing the C fraction, C {\sc ii} starts to develop and 
suppress the emission component of Si {\sc ii} (i.e., the red 
shoulder of Si {\sc ii}). 
If X(C) is 10\%, this effect is clearly seen in the model but not 
in the observed spectra, and thus we can place a conservative upper 
limit of $\sim 10$\% for the carbon fraction at 
$\sim 20,000$ km s$^{-1}$. 
Indeed, the effect is already distinguishable with X(C) $\sim 4$\%, 
which could also be regarded as the upper limit.

In summary, it is inferred that the outermost layers of 
SN 2019ein are structured as follows: (1) the O-Ne-C burning covers the 
region at least between $17,000$ and $30,000$ km s$^{-1}$. 
The inner boundary can exist even deeper. 
(2) If the unburnt layer exists, its inner boundary is 
at least at $30,000$ km s$^{-1}$. 
(3) The mass fraction of unburnt carbon at $\sim 20,000$ km s$^{-1}$ 
is at most 4\% (or 10\% even as a conservative limit). 
(4) There is no mixing of deeper regions processed through more 
advanced nuclear burning out to the O-Ne-C burning/unburnt regions 
studied here. 

\subsection{Implications for Explosion Mechanism and SN Ia Diversity}

The outermost layer of SN 2019ein can be well explained by the 
characteristic composition structure in the O-Ne-C burning layer 
found in the delayed detonation model. 
The (1D) delayed detonation model indeed predicts the 
velocity of this layer similar to that constrained for SN 2019ein; 
$\sim 15,000 - 27000$ km s$^{-1}$, $\sim 17,000 - 32,000$ km s$^{-1}$, 
and $\sim 19,000 - 35,000$ km s$^{-1}$ for the models CS15DD1, 
CS15DD2, and CS15DD3, respectively, in the model sequence of 
\citet{Iwamoto99}. 
The explosion mechanism of SN 2019ein is therefore well represented 
by the delayed detonation model. 
The non-existence of the unburnt region up to at least 
$30,000$ km s$^{-1}$ is also consistent with the delayed 
detonation model.
On the other hand, the same nucleosynthetic layer is confined in a 
small velocity range in the (1D) pure-deflaglation model W7 of 
\citet{Nomoto84}; $\sim 13,000 - 15,000$ km s$^{-1}$, with the 
unbunt later at $> 15,000$ km s$^{-1}$. 

Indeed, in the pioneering study of the `SN spectroscopic tomography' 
by \citet{Stehle05}, they derived similar composition pattern at 
$\sim 16,000 - 23,000$ km s$^{-1}$ for the prototypical 
HV SN 2002bo. 
They introduced a layer of more (slightly) advanced burning stage in 
$\sim 23,000 - 27,000$ km s$^{-1}$, but this might not be robust; 
the first spectrum they modeled had the photospheric velocity 
decreasing to $\sim 16,000$ km s$^{-1}$. 
For SN 2019ein, we do not have to introduce such a composition 
inversion up to $\sim 30,000$ km s$^{-1}$. 
Given the uncertainty involved, we regard that the structure of the 
outermost layer of SN 2019ein shares the similarity to SN 2002bo. 
This might therefore be a common property of HV SNe Ia. 

NV SNe seem to have different characteristics in the 
outermost layer. 
For the best-studied NV SN 2011fe, the composition structure 
derived by \citet{Mazzali14} shows that the O-Ne-C burning region 
is confined in the region at $\sim 13,300 - 19,400$ km s$^{-1}$ 
(with a slight mixing of more advanced burning products including 
$^{56}$Ni, which is not found for SN 2019ein). 
The region at $> 19,400$ km s$^{-1}$ is specified as an unburnt 
layer almost exclusively composed of carbon with primordial metals.
By comparing their findings to those found for SN 2019ein in this 
work, we conclude that SN 2019ein has more extended distribution of 
the O-Ne-C burning layer at higher velocities than SN 2011fe. 

We note that SN 2019ein has larger $\Delta m_{15}$($B$) than 
SN 2011fe, and it is supposed to be fainter with a smaller 
amount of $^{56}$Ni synthesized at the explosion. 
Therefore, the amount of materials processed by the most advanced 
burning (i.e., the complete Si burning) is seemingly smaller for 
SN 2019ein, despite the more extended distribution of the O-Ne-C 
burning region. 
This raises a challenge to the explosion mechanism, 
since there is generally a correlation between the extent of the 
$^{56}$Ni-rich region (or, the mass of $^{56}$Ni) and that of the 
O-Ne-C burning region in the explosion simulations.
For example, this is clearly seen in the delayed detonation model 
sequence of CS15DD1-3 by \citet{Iwamoto99} where the larger 
transition density leads to the more extended regions both for the 
complete Si burning and O-Ne-C burning.

In the above discussion, we have mainly focused on the 
comparison between the properties of SN 2019ein and those expected by 
the delayed detonation model. 
Another explosion model which could also be consistent with our 
constraints is a pure-detonation model of a sub-Chandrasekhar mass WD. 
For example, such a model sequence by \citet{Sim10} suggests 
that the region at $\sim$ 17,000 km s$^{-1}$  is in the O-Ne-C 
burning region which further extends to $> 20,000$ km s$^{-1}$, 
if the progenitor WD mass is below $\sim$ 1.1 M$_{\odot}$. 
However, at the same time, the sub-Chandrasekhar WD explosion may be 
contaminated by the iron-peak elements at the highest velocity, 
in the scenario in which the thermonuclear runaway of the 
sub-Chandrasekhar mass WD is triggered by the surface He 
detonation (i.e., a standard model for the sub-Chandrasekhar 
mass WD explosion; \citealt{Fink10}); such a model would not 
satisfy our constraints placed by the present study.

\section{Conclusions}\label{sec:conclusion}

In this paper, we present photometric and spectroscopic  
observations of HV SN Ia 2019ein starting at 0.3 days 
(photometry) and 2.2 days (spectroscopy) since the discovery. 
We estimate the explosion date as MJD $58602.87 \pm 0.55$, 
i.e., 1.6 days before the discovery, by fitting the expanding 
fireball model to the early rising multi-band light curves. 
This places SN 2019ein as one of the best-studied SNe with the 
observational data starting within a few days since the explosion. 

The light curve evolution shows that SN 2019ein is a relatively 
fast decliner with $\Delta m{15} (B) = 1.36 \pm 0.02$ mag. 
The spectral indicators, i.e., the depth and relative pEW of 
Si {\sc ii} $\lambda$6355 and Si {\sc ii} $\lambda$5972, are 
consistent with $\Delta m_{15} (B)$. 
From the velocity of Si {\sc ii} $\lambda$6355 and its evolution, 
it is robustly classified as a HV SN. 
In summary, SN 2019ein is a HV SN with relatively slow decline; 
this is a previously unexplored type of SNe Ia in terms of the 
observational properties starting within the first few days.

The earliest light curves are used to constrain the radius of 
a possible companion star for SN 2019ein. 
Similarly to other NV SN examples, we do not detect excessive 
emission expected from a giant companion. 
Indeed, our multi-band light curves are well fitted by a single 
power law with the index of 2 (i.e., the fireball model). 

The Si {\sc ii} velocity around the maximum light is modest 
as a HV SN; it is $\sim 13,000$ km s$^{-1}$. 
However, SN 2019ein shows the most rapid decrease in the 
Si {\sc ii} velocity toward the maximum light among 
well-studied SNe Ia including both HV and NV SNe. 
It evolves more rapidly than those HV SNe showing the higher 
maximum-light velocity. 
It is most likely related to the rapid evolution 
in the light curves (i.e., large $\Delta m_{15} (B)$). 
Namely, while it has been reported that the speed of the velocity 
decrease is correlated with the Si {\sc ii} velocity around the 
maximum light, this is not the whole story; the pre-maximum 
velocity evolution does depend on $\Delta m_{15} (B)$, not only 
on the velocity itself.
Therefore, the velocity evolution of 
HV SNe does not form a one-parameter family. 

This additional diversity may be further supported by the behavior 
in the early rising light curves. 
The rise time is suggested to be related with $\Delta m_{15} (B)$. 
HV SNe show generally shorter rise time than NV SNe for given 
$\Delta m_{15} (B)$. 
The short rise time of SN 2019ein fits into the relation. 
Indeed, HV SN 2002er with similar $\Delta m_{15}$ with SN 2019ein 
showed slow rise in its rise, opposite to this relation. 
We note that the Si {\sc ii} velocity of SN 2002er around the 
maximum light is lower than $12,000$ km s$^{-1}$, which is among 
the lowest values to be classified as a HV SN. 
As such, it suggests that the velocity is indeed related to the 
light curve behavior in the rising part. 

We also provide spectral synthesis models for the earliest spectra 
taken within our program (3.7 and 5.7 days since the estimated 
explosion date). 
The phases are sufficiently early to place robust constraints on 
the nature of the outermost layer. 
The layer beyond $\sim 17,000$ km s$^{-1}$ is well described as 
the characteristic O-Ne-C burning region found in the standard 
(1D) delayed detonation model. 
Indeed, we do not find any evidence for mixing of more advance 
burning products from the deeper region. 
This region is extended to at least 25,000 km s$^{-1}$ (likely 
up to $\sim 30,000$ km s$^{-1}$), and there is no unburnt C+O 
material below $\sim 30,000$ km s$^{-1}$. 
This structure is similar to that derived for HV SN 2002bo, 
indicating that the basic structure of the outermost eject of 
HV SNe is not dependent on $\Delta m_{15} (B)$. 
The derived structure is very different from the structure of 
the outermost layer derived for the well-studied NV SN 2011fe, 
where the O-Ne-C burning region is found at much lower velocities 
and confined in a small velocity space. 
Given that the amount of material processed by more advance 
burning (e.g., $^{56}$Ni) should be larger for SN 2011fe than 
SN 2019ein, this might raise a challenge to the 
explosion mechanism. 
We suggest that the relation between the mass of $^{56}$Ni (or 
$\Delta m_{15}$) and the extent of the O-Ne-C provides an important 
constraint on the explosion mechanism(s) of HV and NV SNe. 

\acknowledgments
This research has made use of the NASA/IPAC Extragalactic Database (NED), which 
is operated by the Jet Propulsion Laboratory, California Institute of Technology, 
under contract with the National Aeronautics and Space Administration.
The spectral data of comparison SNe are downloaded from 
SUSPECT\footnote{http://www.nhn.ou.edu/\~{}suspect/} (\citealt{Richardson01}) and 
WISeREP\footnote{http://wiserep.weizmann.ac.il/} (\citealt{Yaron12}) databases.
This research has made use of data obtained from the High Energy Astrophysics 
Science Archive Research Center (HEASARC), a service of the Astrophysics Science 
Division at NASA/GSFC and of the Smithsonian Astrophysical Observatory's High 
Energy Astrophysics Division.
This research made use of \textsc{Tardis}, a community-developed software
package for spectral synthesis in supernovae
\citep{Kerzendorf14, kerzendorf_wolfgang_2019_2590539}.
The development of \textsc{Tardis} received support from the
Google Summer of Code initiative
and from ESA's Summer of Code in Space program. \textsc{Tardis} makes
extensive use of Astropy and PyNE.
This work is supported by the Optical and Near-infrared Astronomy 
Inter-University Cooperation Program.
Part of this work was financially supported by Grants-in-Aid for Scientific 
Research 17H06362 from the Ministry of Education, Culture, Sports, Science 
and Technology (MEXT) of Japan.
This work was supported by the joint research program of the Institute for 
Cosmic Ray Research (ICRR).
M.K. acknowledges support by JSPS KAKENHI Grant (19K23461). 
K.M. acknowledges support by JSPS KAKENHI Grant (18H04585, 18H05223, 
17H02864).
M.Y. is partly supported by JSPS KAKENHI Grant (17K14253).
U.B. acknowledges the support provided by the Turkish 
Scientific and Technical Research Council (T\"{u}B\.{I}TAK-2211C and 
T\"{u}B\.{I}TAK-2214A).


\bibliographystyle{apj}
\bibliography{main}

\end{document}